\documentclass[12pt, hyper]{article}
\usepackage{jheppub1,amsmath,amssymb,amsfonts,amsxtra, mathrsfs, makeidx,graphics,graphicx,amsthm,epsfig, youngtab,bm}

\DeclareMathSizes{12}{10}{7}{7}

\newcommand{\be}{\begin{equation}}
\newcommand{\ee}{\end{equation}}
\newcommand{\beq}{\begin{equation}}
\newcommand{\eeq}{\end{equation}}
\newcommand{\ba}{\begin{array}}
\newcommand{\ea}{\end{array}}
\newcommand{\bi}{\begin{itemize}}
\newcommand{\ei}{\end{itemize}}
\def\vec#1{\bm{#1}}
\def\bea#1\eea{\allowdisplaybreaks \begin{align}#1\end{align}}
 \newcommand{\ben}{\begin{enumerate}}
\newcommand{\een}{\end{enumerate}}
\newcommand{\bean}{\begin{eqnarray*}}
\newcommand{\eean}{\end{eqnarray*}}
\newcommand{\eref}[1]{(\ref{#1})}
\newcommand{\sref}[1]{\S\ref{#1}}
\newcommand{\tref}[1]{Table~\ref{#1}}
\newcommand{\fref}[1]{Figure~\ref{#1}}
\newcommand{\nn}{\nonumber}

\newcommand{\tr}{\mathrm{Tr}}
\newcommand{\PE}{\mathop{\rm PE}}
\newcommand{\PL}{\mathop{\rm PL}}
\newcommand{\tQ}{\widetilde{Q}}

\newcommand{\BC}{\mathbb{C}}
\newcommand{\BR}{\mathbb{R}}

\newcommand{\BZ}{\mathbb{Z}}

\newcommand{\BU}{\mathbf{1}}

\newcommand{\comment}[1]{}

\newcommand{\CM}{{\cal M}}

\newcommand{\CN}{{\cal N}}

\newcommand{\CC}{{\cal C}}

\newcommand{\diag}{\mathrm{diag}}

\newcommand{\Adj}{\mathbf{Adj}}
\newcommand{\fun}{\mathbf{fund}}

\newcommand{\ie}{{\it i.e.}}
\newcommand{\eg}{{\it e.g.}}

\newcommand{\ud}{\mathrm{d}}

\newcommand{\Sym}{\mathrm{Sym}}

\newcommand{\cN}{\mathcal{N}}
\newcommand{\ff}{\mathcal{F}^\flat}

\title{Hilbert Series for Moduli Spaces of Two Instantons}

\author[a]{Amihay Hanany,}
\author[b]{Noppadol Mekareeya,}
\author[c]{and Shlomo S. Razamat}

\affiliation[a]{Theoretical Physics Group, Imperial College London, \\
Prince Consort Road, London,  SW7 2AZ,  UK}
\affiliation[b]{Max-Planck-Institut f\"ur Physik (Werner-Heisenberg-Institut), \\
F\"ohringer Ring 6, 80805 M\"unchen, Deutschland}
\affiliation[c]{Institute for Advanced Study, Princeton, NJ 08540, USA}
\emailAdd{a.hanany@imperial.ac.uk}
\emailAdd{noppadol@mpp.mpg.de}
\emailAdd{razamat@ias.edu}
\abstract{The Hilbert Series (HS) of the moduli space of two $G$ instantons on $\BC^2$, where $G$ is a simple gauge group, is studied in detail. For a given $G$, the moduli space is a singular hyperK\"ahler cone with a symmetry group $U(2) \times G$, where $U(2)$ is the natural symmetry group of $\BC^2$. Holomorphic functions on the moduli space transform in irreducible representations of the symmetry group and hence the Hilbert series admits a character expansion. For cases that $G$ is a classical group (of type $A$, $B$, $C$, or $D$), there is an ADHM construction which allows us to compute the HS explicitly using a contour integral. For cases that $G$ is of $E$-type, recent index results allow for an explicit computation of the HS.  The character expansion can be expressed as an infinite sum which lives on a Cartesian lattice that is generated by a small number of representations. This structure persists for all $G$ and allows for an explicit expressions of the HS to all simple groups. For cases that $G$ is of type $G_2$ or $F_4$, discrete symmetries are enough to evaluate the HS exactly, even though neither ADHM construction nor index is known for these cases.}

\begin{document}
\maketitle

\section{Introduction}
The study of instantons in Yang-Mills theory has received a lot of interest since their discovery in 1975 \cite{Belavin:1975fg,'tHooft:1976fv} and has become a classic subject in theoretical physics.  An important aspect of this subject is to study the space of solutions to the self-dual Yang-Mills equation, known as the {\it moduli space of instantons}.  Such a space has a number of interesting geometrical properties, \eg~  the space is a singular hyperK\"ahler cone. 

For a Yang-Mills theory with a classical gauge group (of types $A$, $B$, $C$ or $D$), a method for constructing the instanton solutions is available and is known as the {\it ADHM construction} \cite{Atiyah:1978ri}.  This construction can be understood from a string theory perspective by considering the system of D$p$ and D$(p+4)$ branes \cite{Witten:1994tz, Douglas:1995bn, Witten:1995gx}.  When D$p$ branes are on top of D$(p+4)$ branes, the former can be realised as instantons moving in four-codimensional worldvolume of the latter.  The gauge theory living on the worldvolume of the D$p$ branes has $8$ supercharges and its Higgs branch can be identified with the corresponding instanton moduli space.  In particular, the $F$ and $D$ term equations give rise to the moment map equations of the corresponding hyperK\"ahler space \cite{AlvarezGaumeFreedman, DeJaegher:1997ka}.  Note that the existence of the ADHM constructions for classical group instantons is closely connected to the existence of the vacuum equations and hence the Lagrangian of the corresponding gauge theory.

The story becomes more complicated and interesting when dealing with exceptional gauge group (of type $E$, $F$ and $G$) instantons.  In such cases, there is no known ADHM construction.\footnote{It should be emphasised that for the $E$-type symmetry, there are several brane constructions, \eg~ using a D4-D8 brane system \cite{Seiberg:1996bd} or using M5 branes on an interval \cite{Ganor:1996mu}.  However, these constructions do not admit a perturbative description since the string coupling is of order 1 and hence there is no Lagrangian.  Note that, using mirror symmetry, one can also realise such an instanton moduli space using the quantum Coulomb branch of the McKay quiver of type $E$ in 3 dimensions.}  
Recently, there have been a proposal that the moduli space of instantons in $E$-type groups can be realised as a Higgs branch of certain 4d $\CN=2$ superconformal theories arising from M5-branes wrapping Riemann surfaces with punctures \cite{Benini:2009gi, Gaiotto:2012uq}.  It is believed that such theories possess no Langrangian description due to the strong coupling and conformality.  Nevertheless, a number of exact quantities, such as indices \cite{Gadde:2009kb,Gadde:2010te, Gadde:2011ik,Gadde:2011uv, Kim:2011mv, Gaiotto:2012uq} and various types of partition functions \cite{Benvenuti:2010pq, Keller:2011ek}, can be computed for instantons in the $E$-type groups.   Note that for the gauge groups of types $F$ and $G$, neither ADHM construction nor such a construction from M5-branes is known; in which case, some other indirect methods have to be applied in order to obtain such exact quantities.

In this paper, we focus on the moduli space of two instantons in an arbitrary simple group on $\BC^2$.  In order to study such a space, we compute an exact quantity that counts chiral gauge invariant operators on the moduli space with respect to a certain $U(1)$ global charge. Such a quantity is known as the {\it Hilbert series}. Hilbert series have been used to study the vacuum structures of a number of supersymmetric gauge theories with various numbers of supersymmetries in various dimensions, regardless of the conformality of theories in question \cite{Pouliot:1998yv, Romelsberger:2005eg, Hanany:2006uc, Benvenuti:2006qr, Feng:2007ur, Forcella:2008bb, Gray:2008yu, Hanany:2008kn, Hanany:2008qc, Davey:2009sr, Benvenuti:2010pq, Hanany:2010qu, Davey:2011mz, Hanany:2012hi}.  It captures both algebraic and geometrical aspects of the moduli space; from which, several pieces of information, such as the generators, the relations and the dimension of the moduli space, can be extracted in a simple way \cite{Benvenuti:2006qr, Feng:2007ur}.  Furthermore, the Hilbert series for the moduli space of $G$ instantons has an interpretation of Nekrasov's partition function for pure super Yang-Mills theory with gauge group $G$ in 5 dimensions \cite{Nakajima:2003pg, Nekrasov:2004vw,Keller:2011ek}.

For a supersymmetric gauge theory with a non-abelian global symmetry $G$, the Hilbert series can be written in terms of infinite sums over characters of representations of $G$; also known as the $G$-invariant {\it character expansion}.   This method allows us to look for expressions that are generic for theories in the same classes; for example 4d $\CN=1$ SQCD with classical gauge groups \cite{Gray:2008yu, Hanany:2008kn}, theories with tri-vertices \cite{Hanany:2010qu} and the moduli space of one instanton in any simple group \cite{Benvenuti:2010pq}.  In this paper, we generalise the results of \cite{Benvenuti:2010pq}, involving the moduli space of one instanton, to the case of two instantons; as can be seen in the main text, the level of complication increases significantly.\footnote{One explanation for such an increasing level of complication is due to a special property of one instanton moduli space. The reduced one $G$ instanton moduli space is known to be the orbit of the highest weight vector in $\mathfrak{g}_\BC$ of $G_\BC$ \cite{Kronheimer} (see also \cite{Gaiotto:2008nz}).  Furthermore, the space of holomorphic functions on such a space is known to be $\oplus_{m=0}^\infty V(m {\vec \alpha}_0)$, where $V(\vec w)$ is the irreducible representation of $G$ with highest weight $\vec w$ and $\vec\alpha_0$ is the highest root of the root system of $G$ (see \eg~ \cite{VinbergPopov,Garfinkle}).  Therefore, one can deduce that the Hilbert series for one instanton moduli space can be written as $\sum_{m=0}^\infty \chi( V(m \vec \alpha_0)) t^{2m}$, where $\chi( V(m \vec \alpha_0))$ denotes the character of the representation whose the highest weight is given by $m$ times that of the adjoint representation of $G$; this agrees with the result in \cite{Benvenuti:2010pq}.  The moduli space of two instantons, however, does not possess this special property.}  Nevertheless, the highest weight vectors of the representations that appear in the character expansion form a lattice whose structure is simple enough to study in a systematic fashion.  This allows us to conjecture and write down the Hilbert series for an arbitrary simple group to all orders in the power expansion.

Another important tool comes from recent developments of ${\cal N}=2$ superconformal index computations.  It was pointed out in~\cite{Gadde:2011uv} that the ${\cal N}=2$  superconformal index in a certain limit of the fugacities  simplifies to a very useful object, dubbed the {\it Hall-Littlewood (HL) index} in~\cite{Gadde:2011uv}.\footnote{
A possible relation of a similar limit of the ${\mathcal N}=1$ index with the
counting problems discussed in~\cite{Gray:2008yu,Hanany:2008kn} was mentioned in~\cite{Spiridonov:2009za}.}
Furthermore, for a theory arising from M5-branes wrapping a Riemann sphere (\ie~ genus 0) with punctures, it is also observed in \cite{Gadde:2011uv} that the HL index is equal to the Hilbert series.  Fortunately, the proposed construction of two instantons in $E$-type gauge groups falls into this category. This allows for the analytic expression of the Hilbert series for instantons in the $E$-type gauge groups; from which, the character expansions can be computed.  

In computing the Hilbert series for two instantons in the gauge groups $F_4$ and $G_2$, we make use of the fact that the Lie algebras of $E_6$ and $SO(8)$ have discrete outer-automorphism groups $\BZ_2$ and $\BZ_3$ respectively.  One can use such discrete symmetries to project the highest weights of $E_6$ and $SO(8)$ representations, respectively, to those of $F_4$ and $G_2$ representations.  We can thus obtain the character expansions for the cases of $F_4$ and $G_2$ in this way.

The paper is organised as follows. 
\bi
 \item In \sref{sec:HSandHL}, we summarise the relations between the HL index and the Hilbert series. Furthermore, in \sref{sec:poletest}, we propose certain properties that should be satisfied by Hilbert series of multi-instantons. These properties are used as a consistency check for the character expansions conjectured in the subsequent sections.
\item The Hilbert series for instantons in classical gauge groups are presented with their ADHM constructions in \S\S \ref{sec:2SUNinsts}--\ref{sec:2SONinsts}.  For each group, we propose a general form of the character expansion valid for generic ranks.  For groups with smaller ranks, such a general form receives some corrections due to irregularities of tensor product decompositions.  We explicitly provide results for such cases.
\item The Hilbert series for instantons in exceptional gauge groups are presented in \S\S \ref{sec:2E6insts}--\ref{sec:2F4insts}.
\item In \sref{sec:universal}, we discuss some universal features of lattices that appear in character expansions for instantons in generic gauge groups.  These include the generators of lattices and the relations between dimensions of lattices and the dimension of the moduli space.
\ei

\subsection{Notation and conventions}  
The following notation and conventions are adopted throughout the paper:
\bi
\item An irreducible representation of a simple group $G$ is denoted by its Dynkin label (or the highest weight) $[a_1, \ldots, a_{\mathrm{rank}~G}]$.  We follow the convention of \href{http://www-math.univ-poitiers.fr/~maavl/LiE/form.html}{LiE online website}\footnote{{\tt http://www-math.univ-poitiers.fr/~maavl/LiE/form.html}}.  A representation of product group $G \times H$ is denoted by $[a_1, \ldots, a_{\mathrm{rank}~G}; b_1, \ldots, b_{\mathrm{rank}~H} ]$, where the representations of $G$ and $H$ are separated by a semi-colon.

\item We indicate the character of a representation using a subscript that corresponds to the symbol used for the fugacity, \eg~$[1,0]_y = y_1 + y_2 y_1^{-1} + y_2^{-1}$.  To avoid cluttered notation, we drop the subscript where there is no potential confusion.

\item Given a simple group $G$, each node in the Dynkin diagram is associated with a simple root of $G$.  In the convention we adopt, the highest weight vector of each simple root is $[0, \ldots, 0, 1, 0, \ldots, 0]$, where the position of $1$ depends on the choice of ordering of the nodes. Here we adopt Bourbaki's convention \cite{bourbaki2002lie}, which is in accordance with the convention adopted by \href{http://www-math.univ-poitiers.fr/~maavl/LiE/form.html}{LiE online website}.

\item In discussing about $k$ $G$ instanton moduli space on $\BC^2$, the moduli space possesses a symmetry $U(2)_{\BC^2} \times G$, where the $U(2)_{\BC^2}$ is the symmetry of $\BC^2$.  We denote the fugacity for the $U(1)$ subgroup of $U(2)_{\BC^2}$ by $t$, the one for the $SU(2)$ subgroup of $U(2)_{\BC^2}$ by $x$, and the ones for $G$ by $\vec y = (y_1, \ldots, y_{\mathrm{rank} G})$.

\item By the {\it reduced} $k$ $G$ instanton moduli space, we mean the moduli space of $k$ $G$ instantons after which the $\BC^2$ component corresponding to the overall position of the instantons has been factored out.   We indicate all quantities associated with the reduced moduli space by tilde, \eg~$\widetilde{\CM}_{k,G}$ denotes the reduced moduli space and $\widetilde{g}_{k,G}$ denotes the corresponding Hilbert series.  The quantity without tilde should be understood as the one associated with the full instanton moduli space; regarding this, note the following relation:
\bea
g_{k,G} (t,x,\vec y) = \frac{1}{1-t x^{\pm1}} \widetilde{g}_{k, G} (t,x, \vec y)~, \qquad 1-t x^{\pm1} \equiv (1-t x)(1-t x^{-1})
\eea
\ei

\section{Hilbert series from Hall-Littlewood indices} \label{sec:HSandHL}
 One way to compute the Hilbert series of the Higgs branch of some ${\cal N}=2$ superconformal theories is through its relation to superconformal index~\cite{Gadde:2011uv}. 
Let us briefly review this relation here.  The superconformal index~\cite{Kinney:2005ej}
 is a partition function of the theory
on $S^3\times S^1$ (with periodic boundary conditions for  fermions around $S^1$).
As such it can be thought of as a trace over the Hilbert space of the theory in radial quantization
which gets contributions only from states anihilated by one of the
supercharges (and its Hermitian 
conjugate).  The states are weighed with fugacities coupled to combinations of charges commuting 
with this  supercharge: these charges are either from the superconformal
algebra or other global ({\it e.g.} flavor) symmetries. The ${\cal N}=2$ superconformal algebra allows for at most three different fugacities of the former type.  It was observed in~\cite{Gadde:2011uv} that by setting  two of the three particularly chosen fugacities to zero the index tremendously 
simplifies: this is the Hall-Littlewood (HL) index. For a theory which can be defined in terms of a Lagrangian the HL index gets contributions
only from one of the complex scalars in the hypermultiplet and one of the fermions in the ${\cal N}=2$ 
vector multiplet.  Thus, this index counts bosonic operators of the Higgs branch 
supplemented with certain fermions coming from the vector fields. On the other hand, the Hilbert 
series of the Higgs branch counts the  same bosonic operators with some operators projected out
by the $F$ term superpotential constraints. 
However, it can be shown~\cite{Gadde:2011uv} that for a class of theories, {\it{e.g.}} linear quivers,
the contribution to the index coming from the fermions matches exactly with 
the projections implied by superpotential constraints, leading to exact equality of the two objects:
the HL index and the Hilbert series. The importance of this equality is that it allows 
for an evaluation of the Hilbert series for theories which are not defined in terms
of a Lagrangian but HL index   of which is known. In particular, the HL index for rank one 
theories with $E$-type flavor symmetry~\cite{Minahan:1996fg,Minahan:1996cj} was computed in~\cite{Gadde:2011uv} and exactly matched with the previous
conjecture for the Hilbert series of their Higgs branch~\cite{Benvenuti:2010pq}.
Here we can make the connection to the problem of this paper: the Higgs branch of the rank
one SCFTs with  $E$-type flavor symmetry is equivalent to the moduli space of one instanton
of $E$-type groups. Similarly,  the Higgs branch of certain higher rank ${\cal N}=2$ SCFTs with 
$E$-type flavor symmetry has been suggested to be equivalent to the moduli space of multi-instantons
of $E$-type groups~\cite{Benini:2009gi,Moore:2011ee}.  The higher rank theories with E-type flavor
symmetry are not defined in terms of Lagrangians but 
explicit expressions for the HL index for them, and thus
equivalently for the Hilbert series of the multi-instanton moduli space of E-type groups,  were constructed in~\cite{Gaiotto:2012uq}. At the moment there are no ${\cal N}=2$ 
superconformal theories Higgs branch of which is suggested to be equivalent to instanton moduli spaces of other exception groups ($G_2$ and $F_4$). 
In this paper we will write the Hilbert series for two instanton moduli space for the classical groups obtained through ADHM techniques and the results for the $E$-type two-instantons obtained through HL index computations in a convenient form, as an infinite sum over characters.
In particular,  this will allow us to suggest analogous expressions for other exceptional cases.

\subsection{Certain properties of the Hilbert series for multi-instantons} \label{sec:poletest}
The moduli space of $k$ $G$ instantons can be approximated by the $k$-th symmetric product of
the moduli space of one instanton. This approximation, of course, has to be corrected by taking into account the 
interaction between the instantons.  However, we propose that certain analytical structures of the 
Hilbert series of both aforementioned spaces remains unchanged under such corrections.
These analytical structures turn out to be useful in checking Hilbert series for two instantons derived in subsequent sections.\footnote{In general, analytical properties of indices are known 
to contain a lot of non-trivial and interesting physical information about theories they 
characterize, see {\it e.g.}~\cite{Gaiotto:2012uq, GRR}.
}

\paragraph{Two $G$ instantons.} Neglecting the interaction between these instantons, we first consider the symmetric square of one $G$ instanton moduli space.  Let $\widetilde{g}_{1,G}(t,\vec y)$ be the Hilbert series of the {\it reduced} one $G$ instanton moduli space.  Note that this does not depend on the $SU(2)_{\BC^2}$ fugacity $x$. The symmetric square ${\text{Sym}^2 \CM_{1,G}}$ of the moduli space $\CM_{1,G}$ of one $G$ instanton, with the overall $\BC^2$ component factored out afterwards, gives rise to the Hilbert series:
\bea \label{sym2}
\widetilde{g}_{\text{Sym}^2 \CM_{1,G}}(t,x, {\bf y}) &= (1-t x^{\pm 1}) \left[ \frac{1}{2} \left \{ \frac{1}{(1-t x^{\pm1})^2} \widetilde{g}^2_{1,G}(t, \vec y) + \frac{1}{1-t^2 x^{\pm2}} \widetilde{g}_{1,G} (t^2, \vec y^2)  \right \} \right] \nn \\
&=   \frac{1}{2} \left[ \frac{1}{(1-t x^{\pm1})} \widetilde{g}^2_{1,G}(t, \vec y) +  \frac{1}{(1+t x^{\pm1})}  \widetilde{g}_{1,G} (t^2, \vec y^2)  \right]~,
%\frac{1}{2} \left(\frac{1}{1-t x^{\pm1}}{\cal I}^2_{\text{(1-inst)}}({\bf y})+\frac{1}{1+t x^{\pm1}} {\cal I}_{\text{(1-inst)}}({\bf y}^2)\right)\,.
\eea 
%where $x$ is the fugacity for $SU(2)_{{\mathbb C}^2}$. 
Note that the two terms on the right-hand-sidehave different meanings. The first one naturally corresponds to the situation in which the two instantons are treated as two distinguishable objects (\ie~ when they are far apart). On the other hand, the second term corresponds to the situation when the two instantons are treated as two identical objects (\ie~ when they are on top of each other and non-interacting). The two terms can be singled out by considering the residues of the Hilbert series at $x=t$ in the former
case and $x=-t$ in the latter case.\footnote{Or equivalently $x^{-1}=t$ and $x^{-1}=-t$.} The Hilbert series in these limits behaves as follows,
\bea\label{deglimits}
 \widetilde{g}_{\text{Sym}^2 \CM_{1,G}}(t,x, {\bf y}) &\sim\frac{1}{x-t}\left[\frac{t}{2}\frac{1}{1-t^2} \widetilde{g}^2_{1,G}(t, \vec y) \right]~, \quad x \rightarrow t \nn \\
 \widetilde{g}_{\text{Sym}^2 \CM_{1,G}}(t,x, {\bf y}) & \sim\frac{1}{x+t}\left[\frac{t}{2}\frac{1}{1-t^2} \widetilde{g}_{1,G} (t^2, \vec y^2) \right]~, \quad x \rightarrow -t ~.
\eea
Now we conjecture that such behaviours in \eref{deglimits} also hold for the reduced moduli space of 2 $G$ instantons.  In other words,
\bea \label{deglimits1}
 \widetilde{g}_{2,G}(t,x, {\bf y}) &\sim\frac{1}{x-t}\left[\frac{t}{2}\frac{1}{1-t^2} \widetilde{g}^2_{1,G}(t, \vec y) \right]~, \quad x \rightarrow t \nn \\
 \widetilde{g}_{2,G}(t,x, {\bf y}) & \sim\frac{1}{x+t}\left[\frac{t}{2}\frac{1}{1-t^2} \widetilde{g}_{1,G} (t^2, \vec y^2) \right]~, \quad x \rightarrow -t ~.
\eea
It can be checked that these conjectures hold for the Hilbert series for two instantons in any classical gauge group (derived from the ADHM constructions) and for the Hall-Littlewood indices for instantons in $E$-type gauge groups (derived in \cite{Gaiotto:2012uq}).\footnote{In fact, this behavior,  
in the limit $x \rightarrow t$,
was first discussed  in~\cite{Gaiotto:2012uq} for the HL index for two $SO(8)$ instantons.}  This leads us to believe that the conjecture should also hold for all simple gauge groups $G$.

In the following subsections, we use \eref{deglimits1} as a consistency check for the character expansion we conjecture for each simple gauge group.  A computationally convenient way to perform such a check on a power series in $t$, order by order, is to use the following equalities:
\bea \label{limxtgen}
(1-t^2) \lim_{x \rightarrow t}~(1- t^2 x^{-2}) \widetilde{g}_{2, G}(t, x, \vec y ) &= \left[ \widetilde{g}_{1, G}(t, \vec y) \right]^2~, \nn \\
(1-t^2) \lim_{x \rightarrow -t}~(1- t^2 x^{-2}) \widetilde{g}_{2, G}(t, x,\vec y ) &=  \widetilde{g}_{1, G}(t^2, \vec y^2) ~.
\eea
Note that these equalities can be easily derived by using \eref{sym2} with the statement of the conjecture.

\paragraph{Higher numbers of instantons.} One can wonder what will be the situation for higher instanton cases. The logic would follow similar lines. The terms in Hilbert series $\widetilde{g}_{\Sym^k \CM_{1,G}} (t, x, \vec y)$ are governed by the partitions of $k$.
As an example, let us consider $k=3$: 
\bea \label{sym3}
 \widetilde{g}_{\Sym^3 \CM_{1,G}} (t, x, \vec y) &= (1-t x^{\pm1}) \left[ \frac{1}{6}  \left \{ g_{1,G} (t, \vec y)^3 +3 g_{1,G} (t, \vec y) g_{1,G} (t^2, \vec y^2)   +2 g_{1,G} (t^3, \vec y^3)  \right \} \right] \nn \\
&= \frac{1}{6}  \Bigg[ \frac{1}{(1-t x^{\pm1})^2} \widetilde{g}_{1,G} (t, \vec y)^3 +3 \frac{1}{1-t^2 x^{\pm2}}  g_{1,G} (t, \vec y) g_{1,G} (t^2, \vec y^2)   \nn \\ 
& \hspace{1cm} +2 \frac{1-t x^{\pm1}}{1-t^3 x^{\pm3}} g_{1,G} (t^3, \vec y^3) \Bigg]~,
\eea
where the terms in the curly bracket come from the cycle index polynomials of the symmetric group $S_3$.  Thus, each term in the expression corresponds, respectively, to the situation in which 
\ben
\item the three instantons are treated as distinguishable objects, \ie~ all of them are far apart from each other,  
\item precisely two of the three instantons are treated as identical objects, \ie~ two of them sit on top of each other (and non-interacting) and the other is far apart,
\item all three instantons are treated as identical objects, \ie~ all of them sit on top of each other and non-interacting.
\een
Observe that each term has the poles at $x=t$, $x=-t$ and $x=\exp(\pm\frac{2\pi i}{3})\,t$.  The behaviour of each term near these poles can be easily  computed by multiplying $\left(1-t^2 x^{-2}\right)\left(1- t^3 x^{-3} \right)$ to both sides of \eref{sym3} and taking the limit $x \rightarrow t$.  

We find that for 3 $SU(2)$ instantons, the Hilbert series $\widetilde{g}_{3,SU(2)}(t, x, \vec y)$ possesses the poles at $x=t$, $x=-t$ and $x=\exp(\pm\frac{2\pi i}{3})\,t$. Moreover, the behaviours of the function near these poles are the same as those of $\widetilde{g}_{\Sym^3 \CM_{1,G}} (t, x, \vec y)$.

\paragraph{General proposal.} In general, we propose that 
\ben
\item Each term in $\widetilde{g}_{\Sym^k \CM_{1,G}} (t, x, \vec y)$, corresponding to the partitions of $k$, possesses poles at positions $x=\exp(i\alpha_\ell)\,t$ with a finite set of values for $\alpha_\ell$.
\item The Hilbert series $\widetilde{g}_{k,G}(t, x, \vec y)$ also possesses poles at positions $x=\exp(i\alpha_\ell)\,t$.  Furthermore, the behaviours of this function as $x \rightarrow \exp(i\alpha_\ell)\,t$ are the same as those of $\widetilde{g}_{\Sym^k \CM_{1,G}} (t, x, \vec y)$.
\een

\section{Two $SU(N)$ instantons} \label{sec:2SUNinsts}
\begin{figure}[htbp]
\begin{center}
\includegraphics[height=0.8 in]{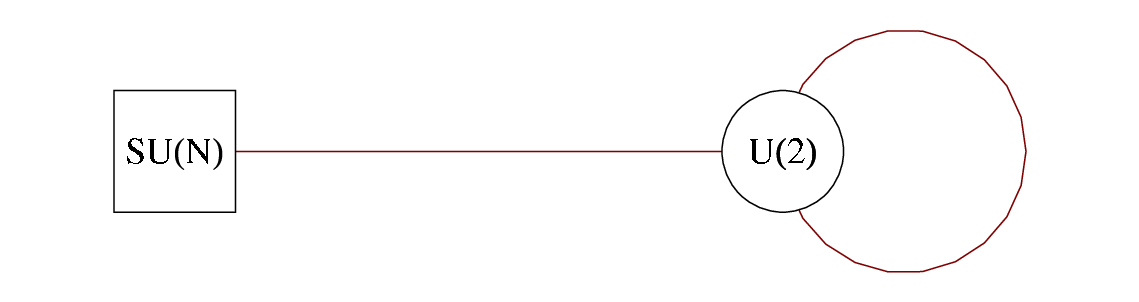}
\caption{The quiver diagram of a 4d $\CN=2$ gauge theory with the gauge group $U(2)$ and a global symmetry $SU(N)$. The matter content consists of a bifundamental hypermultiplet of $U(2) \times SU(N)$, and an adjoint hypermultiplet of the $U(2)$ gauge group. }
\label{fig:N22SU2inst}
\end{center}
\end{figure}

The ADHM data are given by a 4d $\CN=2$ gauge theory whose quiver diagram is depicted in \fref{fig:N22SU2inst}.  We focus on the Higgs branch of this theory.  The moment map equations, which consists of $F$ and $D$ terms in 4d $\CN=1$ language, give rise to a hyperK\"ahler quotient of the moduli space of two $SU(N)$ instantons on $\BR^4$.

Let us emphasise that although we use techniques in four dimensions to study the moduli space, the result still holds in any dimension from 3 to 6. The reason is that hypermultiplet moduli spaces do not depend on the dimension.

\begin{figure}[htbp]
\begin{center}
\includegraphics[height=1.4 in]{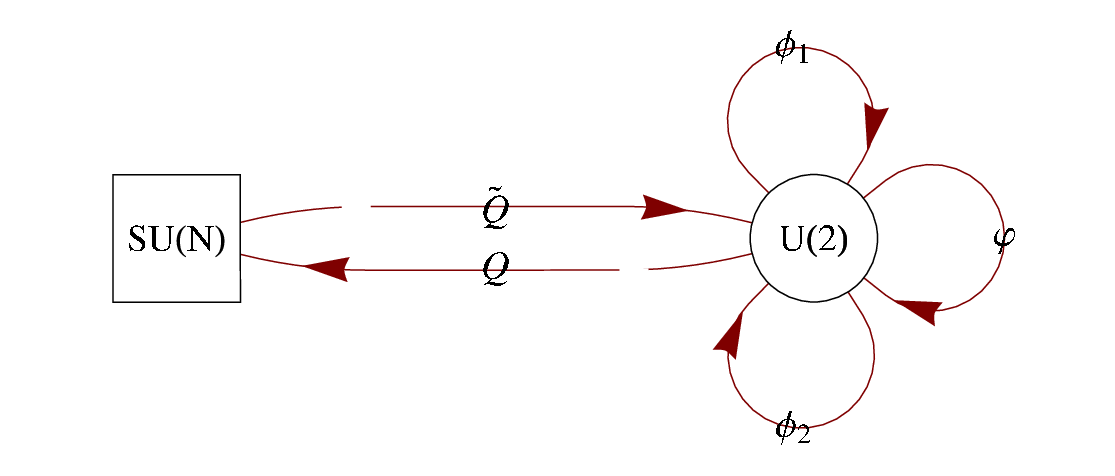}
\caption{The quiver diagram for the theory described by \fref{fig:N22SU2inst}, written in $\cN=1$ notation. The superpotential (setting mass terms to zero) is $W =\tQ_i \cdot \varphi \cdot Q^{i} + \epsilon^{\alpha \beta}  \phi_\alpha \cdot \varphi \cdot \phi_\beta$.}
\label{fig:N12SU2inst}
\end{center}
\end{figure}

In order to compute the Hilbert series, we translate the quiver diagram in \fref{fig:N22SU2inst} to $\cN=1$ language (see \eg~ \cite{Benvenuti:2010pq}).  The corresponding $\cN=1$ quiver diagram is given in \fref{fig:N12SU2inst}.  The field $\varphi$ comes from the scalar in the $\cN=2$ vector multiplet, and the chiral fields $\phi_1, \phi_2, Q, \tQ$ comes from the $\cN=2$ hypermultiplets. 

We have a global symmetry $U(2)_{\BC^2} \times SU(N)$, where $U(2)_{\BC^2}$ corresponds to the isometry of $\BC^2$ parametrised by the overall position of the instantons, and $SU(N)$ corresponds to the square (flavour) node in the quiver diagram.  Note that the global symmetry $U(2)_{\BC^2}$ can be written as $U(1)_{\BC^2} \times SU(2)_{\BC^2}$.    The $U(2)$ adjoint fields $\phi_\alpha = (\phi_1, \phi_2)$ transform as a doublet under $SU(2)_{\BC^2}$ and both $\phi_1$ and $\phi_2$ carry charge $+1$ under $U(1)_{\BC^2}$.   The fields $Q^i_{~a}$ and $Q^a_{~i}$ (with $a = 1,2$ and $i=1,2$) transform under the bi-fundamental representations of $U(2) \times SU(N)$; they transform as a singlet under $SU(2)_{\BC^2}$ and carry charge $+1$ under $U(1)_{\BC^2}$.   We refer the reader to Table 2 of \cite{Benvenuti:2010pq} for more details.

Here and in the rest of the discussion, we use the indices $a,b,c=1,2$ for the gauge symmetry $U(2)$, $i, j, k=1, \ldots, N$ for the global symmetry $SU(N)$, and $\alpha, \beta=1,2$ for the global symmetry $SU(2)_{\BC^2}$.
Due to $\CN=2$ supersymmetry, the superpotential is fixed to be (for simplicity, we set the mass terms to zero):
\bea \label{sup2su2}
W = \tQ_i \cdot \varphi \cdot Q^{i} + \epsilon^{\alpha \beta}  \phi_\alpha \cdot \varphi \cdot \phi_\beta~,
\eea
where `$\cdot$' denotes the contraction of the gauge indices. 

On the Higgs branch, the vacuum expectation values of $\varphi^a_{~b}$ are zero.  Therefore, the $F$-terms are
\bea
0 = \partial_{\varphi^a_{~b}} W = Q^i_{~a} \tQ^b_{~i}  + [\phi_1, \phi_2]^b_{~a}~.
\eea
This matrix equation transforms under the adjoint representation of $U(2)$.  
Therefore, we can write down the Hilbert series of the space of the $F$-term solutions (also known as the $F$-flat space, $\ff$):\footnote{The {\bf plethystic exponential} $\PE$ of a multi-variable function $f(t_1, . . . , t_n)$ that vanishes at the origin, $f(0,...,0) = 0$, is defined as $\PE \left[ f(t_1, t_2, \ldots, t_n) \right] = \exp \left( \sum_{k=1}^\infty \frac{1}{k} f(t_1^k, \ldots, t_n^k) \right)$.} 
\bea
g^{\ff} (t,x,y,z_1,z_2) &= \frac{\PE\left[ \left( z_1+z_2 \right)  [0, \ldots,0,1]_y t +\left( \frac{1}{z_1}+\frac{1}{z_2} \right) [1,0, \ldots,0]_y t +  \left( z_1 + z_2 \right)\left(\frac{1}{z_1}+\frac{1}{z_2} \right) [1]_x t \right]}{\PE \left[ \left( z_1 + z_2 \right)\left(\frac{1}{z_1}+\frac{1}{z_2} \right) t^2   \right]} \nn \\
&= \frac{\prod_{1 \leq a, b \leq 2}  \left(1- \frac{z_a}{z_b} t^2 \right) }{ \left[ \prod_{i = 0}^N \prod_{a=1}^2  \left(1- t z_a \frac{y_i}{y_{i+1}}  \right) \left(1- t z_a ^{-1}\frac{y_{i+1}}{y_{i}}  \right)  \right] \left[ \prod_{\delta = \pm 1} \prod_{1 \leq a, b \leq 2}  \left(1- t \frac{z_a}{z_b} x^\delta  \right)  \right]} \nn \\
& \qquad \text{(with $y_0=y_{N+1} =1$)}~, \label{FF2SUN}
\eea
where $t$ is a fugacity of $U(1)_{\BC^2}$, $x$ is a fugacity of $SU(2)_{\BC^2}$ and $y_1, \ldots, y_{N-1}$ are fugacities of $SU(N)$.  Here $[1]_x = x+ x^{-1}$ is the character of the fundamental representation of $SU(2)_{\BC^2}$, $z_1+z_2$ is the character of the fundamental representation of $U(2)$, and $z_1^{-1}+z_2^{-1}$ is the character of the conjugate representation of the latter. 

The Hilbert series of the Higgs branch of the quiver gauge theory depicted in \fref{fig:N22SU2inst} is given by.
\bea
g_{2,SU(2)}(t,x,y) &= \int \ud \mu_{U(2)} (z_1,z_2)~ g^{\ff} (t,x,y,z_1,z_2) \nn \\
&= \frac{1}{2} \oint_{|z_1| =1} \frac{\ud z_1}{z_1} \oint_{|z_2| =1} \frac{\ud z_1}{z_2} \left(\frac{1}{z_1}-\frac{1}{z_2} \right)(z_1-z_2) \times \nn \\
& \quad \frac{\prod_{1 \leq a, b \leq 2}  \left(1- \frac{z_a}{z_b} t^2 \right) }{ \left[ \prod_{i = 0}^N \prod_{a=1}^2  \left(1- t z_a \frac{y_i}{y_{i+1}}  \right) \left(1- t z_a ^{-1}\frac{y_{i+1}}{y_{i}}  \right)  \right] \left[ \prod_{\delta = \pm 1} \prod_{1 \leq a, b \leq 2}  \left(1- t \frac{z_a}{z_b} x^\delta  \right)  \right]} \nn \\
& \quad \text{(with $y_0=y_{N+1} =1$)}~,
\label{HS2SUNint}
\eea
where the Haar measure of $U(2)$ is given by
\bea \label{HaarU2}
\int \ud \mu_{U(2)} (z_1,z_2) = \frac{1}{2} \oint_{|z_1| =1} \frac{\ud z_1}{z_1} \oint_{|z_2| =1} \frac{\ud z_1}{z_2} \left(\frac{1}{z_1}-\frac{1}{z_2} \right)(z_1-z_2) ~.
\eea

One can compute the integrals \eref{HS2SUNint} using the residue theorem.  It was pointed out in  \cite{Nekrasov:2002qd, Bruzzo:2002xf, Nakajima:2003pg} that the structure of the poles is captured in certain colour partitions of the Young diagrams.  In particular, for $k$ $SU(N)$ instantons, the contributions come from $N$-colour partition of $k$ boxes.  Let us demonstrate the computation for two $SU(2)$ instantons below.

\subsection{Example: Two $SU(2)$ instantons}
For two $SU(2)$ instantons, the Hilbert series can be written as
\bea
g_{2,SU(2)}(t,x,y)
&= \frac{1}{2} \oint_{|z_1| =1} \frac{\ud z_1}{z_1} \oint_{|z_2| =1} \frac{\ud z_1}{z_2} \left(\frac{1}{z_1}-\frac{1}{z_2} \right)(z_1-z_2) \times \nn \\
& \frac{\prod_{1 \leq a, b \leq 2}  \left(1- \frac{z_a}{z_b} t^2 \right) }{\left[ \prod_{\delta_1, \delta_2 = \pm 1} \prod_{a=1}^2  (1- t z^{\delta_1}_a y^{\delta_2}) \right] \left[ \prod_{\delta = \pm 1} \prod_{1 \leq a, b \leq 2}  \left(1- t \frac{z_a}{z_b} x^\delta  \right)  \right]}~. \label{HS2SU2int}
\eea
The integrals in \eref{HS2SU2int} can be computed by summing over the contributions labelled by two-colour partitions of Young diagrams with 2 boxes, namely 
\bea
P_1 = (\tiny{\yng(2)},\cdot),\qquad P_2 = ({\tiny\yng(1,1)},\cdot)~, \qquad P_3 = ({\tiny\yng(1)},{\tiny\yng(1)}),\qquad P_4 = (\cdot, {\tiny\yng(1,1)}), \qquad P_5 = (\cdot, {\tiny\yng(2)})~.
\eea
Let us denote the contribution from $P_k$ by $H_k (t,x,y)$.  Then, the Hilbert series
\bea
g_{2,SU(2)}(t,x,y) = \sum_{k=1}^5 H_k (t,x,y)~, \label{HS2SU2sum}
\eea
where, from Theorem 2.11 of \cite{Nakajima:2003pg}, we have
\bea 
H_1(t,x,y) &= \PE \left[ \frac{1}{x^2}+t \left(\frac{1}{x}+x\right)+t^2 \left(x^2+\frac{1}{y^2}\right)+\frac{t^3 x}{y^2}+y^2+\frac{y^2}{t x} \right]~, \\
H_3(t,x,y) &= \PE \left[ [1]_x (1+ [2]_y) t \right]~,
\eea
and the remaining $H_k$ are determined by the following relations:
\bea
\begin{array}{ll}
H_1 (t,x,y) = H_2 (t,1/x,y)~, \qquad & H_4 (t,x,y) = H_5 (t,1/x,y)~, \\
H_1 (t,x,y) = H_5(t,x,1/y)~, \qquad & H_2(t,x,y) = H_4(t,x, 1/y)~.
\end{array}
\eea

Performing the summation in \eref{HS2SU2sum}, we obtain the Hilbert series of the moduli space of 2 $SU(2)$ instantons on $\BR^4$:
\bea
g_{2,SU(2)} (t,x,y) = \widetilde{g}_{2,SU(2)} (t,x,y) \PE[[1]_x t]~. \label{includeCoM2SU2}
\eea
where $\widetilde{g}_{2,SU(2)} (t,x,y)$ and $\PE[[1]_x t]$ are given as follows:
\bi
\item The Hilbert series of $\BC^2$ parametrised by the overall position of the instantons is
\bea
\PE[[1]_x t] = \frac{1}{(1- t x)(1-t/x)}~, 
\eea
\item The function $\widetilde{g}_{2,SU(2)} (t,x,y)$ has an interpretation as the Hilbert series of the {\bf reduced instanton moduli space} (\ie~neglecting the overall position).  It admits the character expansion in terms of representations of $SU(2)_{\BC^2} \times SU(2)$ and can be written as
\bea
 \widetilde{g}_{2,SU(2)} (t,x,y)  &=  \frac{1}{1-t^4} \sum_{m_2=0}^\infty \sum_{n_2=0}^\infty \sum_{n_3=0}^\infty   \Big \{ [2m_2+n_3; 2n_2+2n_3]_{x;y}~t^{2m_2+2n_2+3n_3}  \nn \\
& +[2m_2+n_3+1; 2n_2+2n_3+2]_{x;y}~t^{2m_2+2n_2+3n_3+5} \Big \}~. \label{Irr2SU2}
\eea
where $[r_1;r_2]$ denotes the representation of the global symmetry $SU(2)_{\BC^2} \times SU(2)$
\ei
%the Hilbert series $\widetilde{g}_{2,SU(2)} (t,x,y)$ of the reduced instanton moduli space admits the character expansion in terms of representations of $SU(2)_{\BC^2} \times SU(2)_F$ and can be written as
%\bea
% \widetilde{g}_{2,SU(2)} (t,x,y)  &=  \frac{1}{1-t^4} \sum_{m_2,n_2,n_3 =0}^\infty  \Big \{ [2m_2+n_3; 2n_2+2n_3]_{x;y}~t^{2m_2+2n_2+3n_3}  \nn \\
%& +[2m_2+n_3+1; 2n_2+2n_3+2]_{x;y}~t^{2m_2+2n_2+3n_3+5} \Big \}~, \label{Irr2SU2}
%\eea
%where $[r_1;r_2]$ denotes the representation of the global symmetry $SU(2)_{\BC^2} \times SU(2)_F$, and
%the Hilbert series of $\BC^2$ parametrised by the overall position of the instantons is given by
%\bea
%\PE[[1]_x t] = \frac{1}{(1- t x)(1-t/x)}~.
%\eea

Let us focus on the Hilbert series of the irreducible component of the instanton moduli space.
The plethystic logarithm of this Hilbert series is given by
\bea
\PL \left[ \widetilde{g}_{2,SU(2)} (t,x,y) \right] &= ([2;0]+[0;2]) t^2 + [1;2] t^3 - t^4 - ([1;2]+[1;0]) t^5 \nn \\
& - ([2;0] +[0;2])t^6 + \ldots ~.
\eea
This implies that the generators of the moduli space are
\bi
\item {\bf Order $t^2$:}  The symmetric traces $S_{\alpha \beta} := \tr (\phi_\alpha \cdot \phi_\beta)$ in the representation $[2;0]$, and the mesons $M^i_{~j} := Q^i_{~a} \tQ^a_{~j}$ (subject to the relation $M^i_{~i} =0$ from the $F$-terms) in the representation $[0;2]$.
\item {\bf Order $t^3$:} The adjoint mesons  $(A_\alpha)^i_{~j} = Q^i \cdot \phi_\alpha \cdot \tQ_j$.  Note that from the $F$-terms, $(A_\alpha)^i_{~i} = \tQ^b_{~i} Q^i_{~a} (\phi_\alpha)^a_{~b} = -\tr \left( [\phi_1,\phi_2] \cdot \phi_\alpha \right) =0$.  Hence, $(A_\alpha)^i_{~j} $ are in the representation $[1;2]$.
\ei
Setting $x=1, y=1$, we obtain the unrefined Hilbert series
\bea
 \widetilde{g}_{2,SU(2)} (t)  = \frac{1+t+3 t^2+6 t^3+8 t^4+6 t^5+8 t^6+6 t^7+3 t^8+t^9+t^{10}}{(1-t)^6 (1+t)^4 \left(1+t+t^2\right)^3}~.
\eea
The order of the pole at $t=1$ indicates that the reduced instanton moduli space $\widetilde{\CM}_{2,SU(2)}$ is 6 complex dimensional or, equivalently, 3 quaternionic dimensional as expected.

%Substituting into \eref{includeCoM2SU2} the following variables:
%\bea
%t = \exp \left( - \frac{1}{2}\beta (\epsilon_1 +\epsilon_2) \right),\quad y = \exp (-\beta a), \quad x = \exp \left( - \frac{1}{2}\beta (\epsilon_1 -\epsilon_2) \right)~,
%\eea
%and considering the limit $\beta \rightarrow 0$, we obtain 
%\bea
%g_{2,SU(2)} (\epsilon_1, \epsilon_2 ,a, \beta)= \beta^{-8} Z^{4d}_{G=2,SU(2)} (\epsilon_1, \epsilon_2 ,a)~, 
%\eea
%where  $Z^{4d}_{G=2,SU(2)} (\epsilon_1, \epsilon_2 ,a)$ is the four dimensional Nekrasov partition function:
%\bea
% Z^{4d}_{G=2,SU(2)} (\epsilon_1, \epsilon_2 ,a) = \frac{1}{\epsilon _1^2 \epsilon _2^2}  \frac{\epsilon_1 \epsilon_2 +8 (\epsilon _1+ \epsilon _2)^2 -8 a^2}{ \left[(\epsilon _1+\epsilon _2)^2- 4a^2\right] \left[(2\epsilon _1+\epsilon _2)^2- 4a^2\right]\left[(\epsilon _1+2\epsilon _2)^2- 4a^2\right] }~.
%\eea

\subsection{General formula}
One way to proceed to higher $SU(N)$ groups is to use either \eref{HS2SUNint} or the method of summing over contributions of coloured partitions proposed by \cite{Nekrasov:2002qd, Bruzzo:2002xf, Nakajima:2003pg}.  These expressions are rather long and there is no clear generalisation of the Hilbert series to higher values of $N$ or other simple groups.  Although there is a proposal  \cite{Hollands:2010xa} to generalise the method of coloured partition for $SU(N)$ instantons to the cases of $SO(N)$ and $Sp(N)$, such generalisation can be rather computationally involved.

Instead we choose to proceed by performing character expansion of the Hilbert series and looking for expressions that are generic for all $SU(N)$ groups, or more generally expressions which allow generalisation to other groups.

In fact, the approach of using character expansion to evaluate Hilbert series to arbitrary order proved to be rather successful and has been applied in various examples \cite{Forcella:2008bb, Gray:2008yu, Hanany:2008kn, Hanany:2008qc, Davey:2009sr, Benvenuti:2010pq, Hanany:2010qu, Davey:2011mz, Hanany:2012hi}. It turns out that this approach is useful for the problem at hand.  To understand this, one observes that the Dynkin labels of irreducible representations live on lattices, and due to the structure of tensor products, which always starts as additive higher order representations are points on a conical lattice which is generated by a very small number of representations.  We proceed by giving the conjecture for the character expansion of the Hilbert series for two $SU(N)$ instantons, followed by explanations on the structure of the lattices.

The Hilbert series for the reduced two $SU(N)$ instanton moduli space is conjectured to be
{\scriptsize
\bea \label{gen2SUN}
& \widetilde{g}_{2,SU(N)} (t,x,y_1, \ldots, y_{N-1})  \nn \\
& =   \sum_{k_4 =0}^\infty   \Bigg \{ f(0; k_4, 0, 0,\ldots,0, 0, k_4) t^{ k_4} +f(1; k_4+1, 0, 0,\ldots,0, 0, k_4+1) t^{ 4k_4+5} \Bigg \} \nn \\
& + \sum_{k_4 =0}^\infty\sum_{k_6 =0}^\infty   \Bigg \{ \Big[ f(0; k_4, k_6 + 1, 0,\ldots,0, 0, k_4+ 2 k_6 + 2) +f(0; k_4+ 2 k_6 + 2, 0, 0,\ldots,0, k_6 + 1, k_4) \Big] t^{4 k_4 + 6 k_6 + 6} \nn \\
&\qquad + \Big[ f(1; k_4, k_6 + 1, 0,\ldots,0, 0, k_4+ 2 k_6 + 2) +f(1; k_4+ 2 k_6 + 2, 0, 0,\ldots,0, k_6 + 1, k_4) \Big] t^{4 k_4 + 6 k_6 + 7} \Bigg \}  \nn \\
& + \sum_{k_5 =0}^\infty\sum_{k_6 =0}^\infty   \Bigg \{ \Big[ f(k_5+1; 0,k_5+k_6+1,0, \ldots,0, 2k_5 +2k_6+2) + f(k_5+1; 2k_5 +2k_6+2,0, \ldots,0, k_5+k_6+1,0) \Big] t^{5 k_5 + 6 k_6 + 5} \nn \\
& \qquad + \Big[ f(k_5+2; 0,k_5+k_6+2,0, \ldots,0, 2k_5 +2k_6+4) + f(k_5+2; 2k_5 +2k_6+4,0, \ldots,0, k_5+k_6+2,0) \Big] t^{5 k_5 + 6 k_6 + 12}   \Bigg \}~,
\eea}
where the function $f$ is defined as follows:
\bea
 f(a;b_1, b_2, \ldots, b_{N-1}) &= \frac{1}{1-t^4} \sum_{m_2=0}^\infty \sum_{n_2=0}^\infty \sum_{n_3=0}^\infty \sum_{n_4=0}^\infty t^{2 m_2+2 n_2+3 n_3+4 n_4} \times \nn \\
&[2m_2+n_3+a; n_2+n_3+b_1, n_4 + b_2, b_3, b_4,\ldots, b_{N-3}, n_4+b_{N-2},  n_2+n_3+b_{N-1}]~. \label{deffSUN}
\eea

There are two sets of the generators of the reduced instanton moduli space of two $SU(N)$ instantons: 
\bi
\item At order $t^2$, the generators transform under the $SU(2) \times SU(N)$ representation 
\bea [2;0,\ldots, 0]+ [0;1,0,\ldots,0,1] = [\Adj_{SU(2)}; \mathbf{singlet}_{SU(N)}]+ [\mathbf{singlet}_{SU(2)}; \Adj_{SU(N)}]~. \eea
\item At order $t^3$, the generators transform under
\bea [1;1,0, \ldots, 0] = [\mathbf{\fun}_{SU(2)}; \Adj_{SU(N)}]~. \eea
\ei
Note that the representation $[\Adj_{SU(2)}; \mathbf{singlet}_{G}]+ [\mathbf{singlet}_{SU(2)}; \Adj_{G}]$ at order $t^2$ and the representation $[\mathbf{\fun}_{SU(2)}; \Adj_{G}]$ at order $t^3$ persist for the generators of two $G$ instanton moduli space, with any simple group $G$.

It should be emphasise that general formula \eref{gen2SUN} takes its exact form for $N\geq5$.   For smaller $N$, the general formula receives some corrections due to irregularities of the highest weight representations in tensor product decompositions.  We discuss this in details in \sref{specialcasesSU}.

\subsubsection{The lattice structure} \label{latticeSU}
Observe that the terms in $f(0; 0, \ldots, 0)$ can be viewed as points in a five-dimensional lattice.  This lattice is spanned by certain highest weight vectors associated with $SU(2)\times SU(N)$ representations.  We can determine three elements in the basis set out of five by looking at the generators of the moduli space.  In particular, at order $t^2$ the generators of the moduli space transform in the $SU(2)\times SU(N)$ representation $[2;0,0,\ldots, 0]+[0;1,0,\ldots,0,1]$, and at order $t^3$ the generators transform in $[1;1,0,\ldots,0,1]$.  The directions spanned by these vectors are denoted by $m_2$, $n_2$ and $n_3$ respectively.  The remaining basis vectors can be obtained by studying the Hilbert series at order $t^4$; as can be seen from \eref{deffSUN}, these basis vectors are $[0;0,0, \ldots,0]$ and $[0;0,1,0, \ldots,0,1,0]$.  The directions spanned by these vectors is indicated by the factor $\frac{1}{1-t^4}$ and the index $n_4$ respectively. 

\begin{figure}[htbp]
\begin{center}
\includegraphics[scale=1]{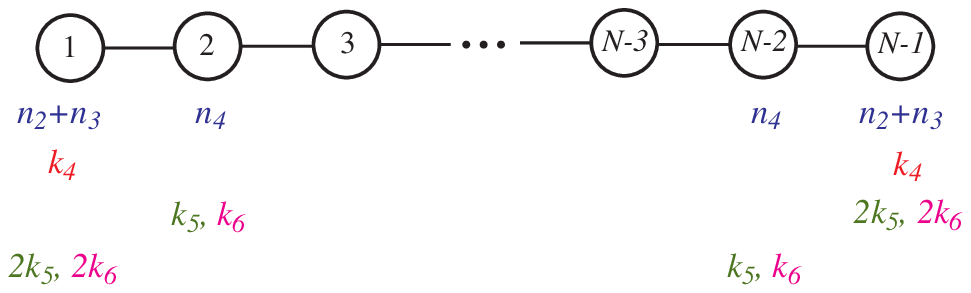}
\caption{The Dynkin diagram of $SU(N)$.  The labels in black indicate ordering of the nodes; the one with number $n$ can be associated with the representation $[0,\ldots, 0, 1, 0, \ldots, 0]$ of $SU(N)$, with $1$ in the $n$-th position from the left.   The labels in blue indicate the indices in the universal lattice $f(0;0,\ldots,0)$.  The labels in red, green and pink indicate the indices in the non-universal lattices.  Observe that the lattices in \eref{gen2SUN} occupy only 4 nodes of the Dynkin diagram in a symmetric fashion, two from each end.}
\label{DynkinAn}
\end{center}
\end{figure}

Note that the $SU(N)$ adjoint representation $[1,0,\ldots,0,1]$ can be associated with the first and the last node of the Dynkin diagram, and the representation $[0,1,0, \ldots,0,1,0]$ can be associated with the second node and the node next to the last.  In \fref{DynkinAn}, we depict the $SU(N)$ Dynkin diagram with the ordering of the nodes; in general, the node with number $n$ can be associated with the representation $[0,\ldots, 0, 1, 0, \ldots, 0]$ of $SU(N)$, with $1$ in the $n$-th position from the left.

General formula \eref{gen2SUN} is written in terms of various summations of the function $f$ evaluated at various points.  Note that the lattice in $f(0;0, \ldots,0)$ appears universally in \eref{gen2SUN}, we refer to such a lattice, together with the corresponding powers of $t$, as the {\bf universal lattice}.  The summands in \eref{gen2SUN} also consist of other lattices associated with the indices $k$'s; in particular, $k_n$ is associated with the generator of such a lattice at order $t^n$.    Since each $k_n$ does not appear universally in the formula \eref{gen2SUN}, we refer to the latter lattices as {\bf non-universal lattices}. In addition to such lattices, there are also vectors that indicates {\bf shifts} from the universal and non-universal lattices, \eg~ at order $t^7$, the shifts are $[1; 0,  1, 0,\ldots,0, 0, 2] +[1;  2, 0, 0,\ldots,0, 1, 0]$.

Since each summation in \eref{gen2SUN} runs from zero to infinity by default, the universal lattice, the non-universal lattices and the shifts fix general formula \eref{gen2SUN} uniquely.  We present the structures of the universal lattice and the non-universal lattices in \fref{DynkinAn}.

One crucial observation is that the lattices in \eref{deffSUN} occupy only 4 positions of the Dynkin label of $SU(N)$ in a symmetric fashion, namely two from the left and two from the right, and the numbers appearing in the remaining positions are identically zero. Such four positions corresponds to four nodes of the Dynkin diagram of $SU(N)$, as shown in \fref{DynkinAn}.

\subsubsection{Testing the conjecture}  
Conjecture \eref{gen2SUN} can be tested in several non-trivial ways.  Let us present two of them.  First, we set $x= y_1 = \ldots, = y_{N-1} =1$; the result must satisfy the following conditions:
\bi
\item The summations yield a rational function in $t$ with a palindromic numerator.  This is because the moduli space is a hyperK\"ahler cone, which is a Calabi-Yau variety.
\item The pole at $t=1$ is of order $4N-2$. This is because the reduced instanton moduli space is $4N-2$ complex dimensional.
\ei
For reference, we write down the unrefined Hilbert series for a few values of $N$ below:
\bea
\widetilde{g}_{2,SU(2)} (t) &= \frac{1+t+3 t^2+6 t^3+8 t^4+6 t^5+8 t^6+6 t^7+3 t^8+t^9+t^{10}}{(1-t)^6 (1+t)^4 \left(1+t+t^2\right)^3}~, \nn \\
\widetilde{g}_{2,SU(3)} (t) &= \frac{1}{(1-t)^{10} (1+t)^6 \left(1+t+t^2\right)^5}\Big(1+t+6 t^2+17 t^3+31 t^4+52 t^5+92 t^6+110 t^7 \nn \\
&\quad +112 t^8+110 t^9+92 t^{10}+52 t^{11}+31 t^{12}+17 t^{13}+6 t^{14}+t^{15}+t^{16}\Big)~, \nn \\
\widetilde{g}_{2,SU(4)} (t) &= \frac{1}{(1-t)^{14} (1+t)^8 (1+t+t^2)^7} \Big(1+t+11 t^2+34 t^3+88 t^4+216 t^5+473 t^6+797 t^7 \nn \\
& \quad +1243 t^8+1738 t^9+2080 t^{10}+2152 t^{11}+2080 t^{12}+1738 t^{13}+1243 t^{14}+797 t^{15} \nn \\
& \quad +473 t^{16}+216 t^{17}+88 t^{18}+34 t^{19}+11 t^{20}+t^{21}+t^{22} \Big)~, \nn \\
\widetilde{g}_{2,SU(5)} (t) &= \frac{1}{(1-t)^{18} (1+t)^{12} (1+t+t^2)^9} \Big(1+3 t+21 t^2+94 t^3+341 t^4+1099 t^5+3137 t^6 \nn \\
& \quad +7624 t^7+16442 t^8+31830 t^9+55082 t^{10}+85360 t^{11}+120008 t^{12}+153060 t^{13}+176628 t^{14} \nn \\
& \quad +184960 t^{15}+176628 t^{16}+153060 t^{17}+120008 t^{18}+85360 t^{19}+55082 t^{20}+31830 t^{21} \nn \\
& \quad +16442 t^{22}+7624 t^{23}+3137 t^{24}+1099 t^{25}+341 t^{26}+94 t^{27}+21 t^{28}+3 t^{29}+t^{30}\Big)~, \nn \\
\widetilde{g}_{2,SU(6)} (t) &= \frac{1}{(1-t)^{22} (1+t)^{16} \left(1+t+t^2\right)^{11}} \Big(1+5 t+37 t^2+204 t^3+947 t^4+3819 t^5+13587 t^6 \nn \\
& \quad +42180 t^7+116511 t^8+289075 t^9+647517 t^{10}+1314730 t^{11}+2435034 t^{12}+4128428 t^{13} \nn \\
& \quad +6422514 t^{14}+9189070 t^{15}+12121994 t^{16}+14760964 t^{17}+16603650 t^{18}+17264534 t^{19} \nn \\
& \quad +16603650 t^{20} + \text{palindrome up to $t^{38}$} \Big)~.
\eea
We observe that for $N\geq4$ the unrefined Hilbert series takes the following form:
\bea
\widetilde{g}_{2,SU(N)} (t) &= \frac{P_{8N-10}(t)}{(1-t)^{4N-2}(1+t)^{4N-8}(1+t+t^2)^{2N-1}}~,
\eea
where $P_{8N-10}(t)$ is a palindromic polynomial of degree $8N-10$.

The second test is to check that \cite{Gaiotto:2012uq} the character expansion for $\widetilde{g}_{2, SU(N)}(t, x, \{ y_i  \})$ satisfies the limits \eref{limxtgen}.  We have performed such a test for $N=2, \ldots, 6$ and the results are as required.
%where $\widetilde{g}_{1, SU(N)}(t, \{ y_i \})$ denotes the Hilbert series of reduced 1 $SU(N)$ instanton moduli space:
%\bea
%\widetilde{g}_{1, SU(N)}(t, \{ y_i \}) = \sum_{n=0}^\infty [n,0, \ldots, 0,n] t^{2n}~.
%\eea
%Note that condition \eref{limxt} implies that
%\bea
%2(1-t^2)t^{-1} \mathrm{Res}_{x \rightarrow t}~\widetilde{g}_{2, SU(N)}(t, x, \{ y_i \} )  = \left[ \widetilde{g}_{1, SU(N)}(t, x=1, \{ y_i \}) \right]^2~.
%\eea
%We use condition \eref{limxt} to test conjecture \eref{gen2SUN} for $N=2, \ldots, 6$.

\subsubsection{Special cases of low rank groups: $SU(2)$, $SU(3)$ and $SU(4)$} \label{specialcasesSU}
For the cases $SU(2)$, $SU(3)$ and $SU(4)$, there are, respectively, only $1$, $2$ and $3$ nodes in the Dynkin diagrams.   Therefore, the lattice structure depicted in \fref{DynkinAn} may not appear fully in such cases, and we thus expect that there are corrections to general formula \eref{gen2SUN} for the cases of $N=2,3,4$.   

\begin{figure}[htbp]
\begin{center}
\includegraphics[scale=0.4]{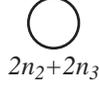}
\caption{The Dynkin diagram of $SU(2)$. The label indicate the indices in the universal lattice.}
\label{DynkinA1}
\end{center}
\end{figure}

\paragraph{The case of $SU(2)$.} The Hilbert series is given by \eref{Irr2SU2}.  The formula contains precisely one four-dimensional lattice generated by $[2;0]+[0;2]$ at order $t^2$, $[1;2]$ at order $t^3$ and $[0;0]$ at order $t^4$.  Note that the former three also correspond to the generators of the moduli space.  The lattice structure can be summarised in \fref{DynkinA1}.

\begin{figure}[htbp]
\begin{center}
\includegraphics[scale=0.4]{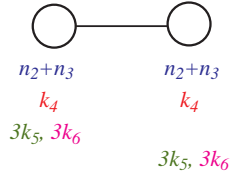}
\caption{The Dynkin diagram of $SU(3)$. The labels in blue indicate the indices in the universal lattice $f(0;0,\ldots,0)$.  The labels in red, green and pink indicate the indices in the non-universal lattices. Note that the projection from \eref{gen2SUN} to \eref{gen2SU3} removes the index $n_4$, keeps $k_4$, and acts additively on $k_5$, $k_6$.}
\label{DynkinA2}
\end{center}
\end{figure}

\paragraph{The case of $SU(3)$.} For the case of $SU(3)$, the Hilbert series can be computed from the ADHM construction. The result is as follows:
\bea \label{gen2SU3}
& \widetilde{g}_{2,SU(3)} (t,x,y_1,y_2)  \nn \\
& =   \sum_{k_4 =0}^\infty   \Bigg \{ f(0; k_4, k_4) t^{ k_4} +f(1; k_4+1, k_4+1) t^{ 4k_4+5} \Bigg \} \nn \\
& + \sum_{k_4 =0}^\infty\sum_{k_6 =0}^\infty   \Bigg \{ \Big[ f(0; k_4 + 3 k_6 + 3, k_4) +f(0; k_4,k_4 + 3 k_6 + 3) \Big] t^{4 k_4 + 6 k_6 + 6} \nn \\
&\qquad \qquad \quad + \Big[ f(1; k_4 + 3 k_6 + 4, k_4+1) +f(1; k_4+1,k_4 + 3 k_6 + 4) \Big] t^{4 k_4 + 6 k_6 + 11} \Bigg \}  \nn \\
& +  \sum_{k_5 =0}^\infty \sum_{k_6 =0}^\infty  \Bigg \{ \Big[ f(k_5+1; 3k_5+3k_6+3, 0) + f(k_5+1;0,3k_5+3k_6+3) \Big] t^{5 k_5 + 6 k_6 + 5} \nn \\
& \qquad \qquad \quad + \Big[ f(k_5+1; 3 k_5 + 3 k_6 + 3,0) + f(k_5+1;0, 3 k_5 + 3 k_6 + 3) \Big] t^{5 k_5 + 6 k_6 + 7}   \Bigg \}~,
\eea
where the function $f$ is defined as follows:
\bea
 f(a;b_1, b_2) &= \frac{1}{1-t^4} \sum_{m_2=0}^\infty \sum_{n_2=0}^\infty \sum_{n_3=0}^\infty [2m_2+n_3+a; n_2+n_3+b_1,  n_2+n_3+b_{2}]  t^{2 m_2+2 n_2+3 n_3}~. \label{deffSU3}
\eea
The lattice structures are summarised in \fref{DynkinA2}. Note that the projection from \eref{gen2SUN} to \eref{gen2SU3} removes the index $n_4$, keeps $k_4$, and acts additively on $k_5$, $k_6$.

\begin{figure}[htbp]
\begin{center}
\includegraphics[scale=0.7]{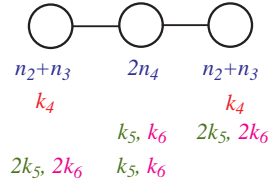}
\caption{The Dynkin diagram of $SU(4)$. The labels in blue indicate the indices in the universal lattice $f(0;0,\ldots,0)$.  The labels in red, green and pink indicate the indices in the non-universal lattices. Note that the projection from \eref{gen2SUN} to \eref{gen2SU4} acts trivially on all indices, except for $n_4$, in which case this is additive.}
\label{DynkinA3}
\end{center}
\end{figure}

\paragraph{The case of $SU(4)$.}  For the case of $SU(4)$, the Hilbert series can be computed from the ADHM construction. The result is as follows:
{\footnotesize
\bea \label{gen2SU4}
& \widetilde{g}_{2,SU(4)} (t,x,y_1,y_2,y_3)  \nn \\
& =   \sum_{k_4 =0}^\infty   \Bigg \{ f(0; k_4,0, k_4) t^{ k_4} +f(1; k_4+1,0, k_4+1) t^{ 4k_4+5} \Bigg \} \nn \\
& + \sum_{k_4 =0}^\infty\sum_{k_6 =0}^\infty   \Bigg \{ \Big[ f(0; k_4,1+k_6,2+k_4+2 k_6) +f(0; 2+k_4+2 k_6,1+k_6,k_4) \Big] t^{4 k_4 + 6 k_6 + 6} \nn \\
&\qquad \qquad \quad + \Big[ f(1; k_4,1+k_6,2+k_4+2 k_6) +f(1; 2+k_4+2 k_6,1+k_6,k_4) \Big] t^{4 k_4 + 6 k_6 + 7} \Bigg \}  \nn \\
& +  \sum_{k_5 =0}^\infty \sum_{k_6 =0}^\infty  \Bigg \{ \Big[ f(k_5+1; 0,1+k_5+k_6,2+2 k_5+2 k_6) + f(k_5+1;2+2 k_5+2 k_6,1+k_5+k_6,0) \Big] t^{5 k_5 + 6 k_6 + 5} \nn \\
& \qquad \qquad \quad + \Big[ f(k_5+2; 0,2+k_5+k_6,4+2 k_5+2 k_6) + f(k_5+2;4+2 k_5+2 k_6,2+k_5+k_6,0) \Big] t^{5 k_5 + 6 k_6 + 12}   \Bigg \}~,
\eea}
where the function $f$ is defined as follows:
\bea
 f(a;b_1, b_2,b_3) &= \frac{1}{1-t^4} \sum_{m_2=0}^\infty \sum_{n_2=0}^\infty \sum_{n_3=0}^\infty [2m_2+n_3+a; n_2+n_3+b_1, 2n_4+b_2,  n_2+n_3+b_{3}]  t^{2 m_2+2 n_2+3 n_3}~. \label{deffSU4}
\eea
The lattice structures are summarised in \fref{DynkinA3}.  Note that the projection from \eref{gen2SUN} to \eref{gen2SU4} acts trivially on all indices, except for $n_4$, in which case this is additive.

\section{Two $Sp(N)$ instantons}
\begin{figure}[htbp]
\begin{center}
\includegraphics[height=0.8 in]{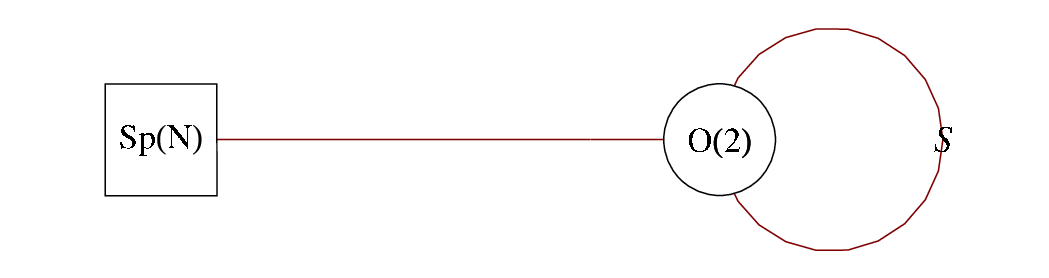}
\caption{The quiver diagram of a 4d $\CN=2$ gauge theory with the gauge group $O(2)$ and a global symmetry $Sp(N)$. The matter content consists of a bifundamental hypermultiplet of $O(2) \times Sp(N)$, and the rank-2 symmetric hypermultiplet $S$ of the $O(2)$ gauge group. }
\label{fig:N22Sp1inst}
\end{center}
\end{figure}

In this section, we compute the Hilbert series of the moduli space of two $Sp(N)$-instantons on $\BR^4$.  The ADHM data are given by the four dimensional $\CN=2$ quiver gauge theory depicted in \fref{fig:N22Sp1inst}.

\begin{figure}[htbp]
\begin{center}
\includegraphics[scale=0.7]{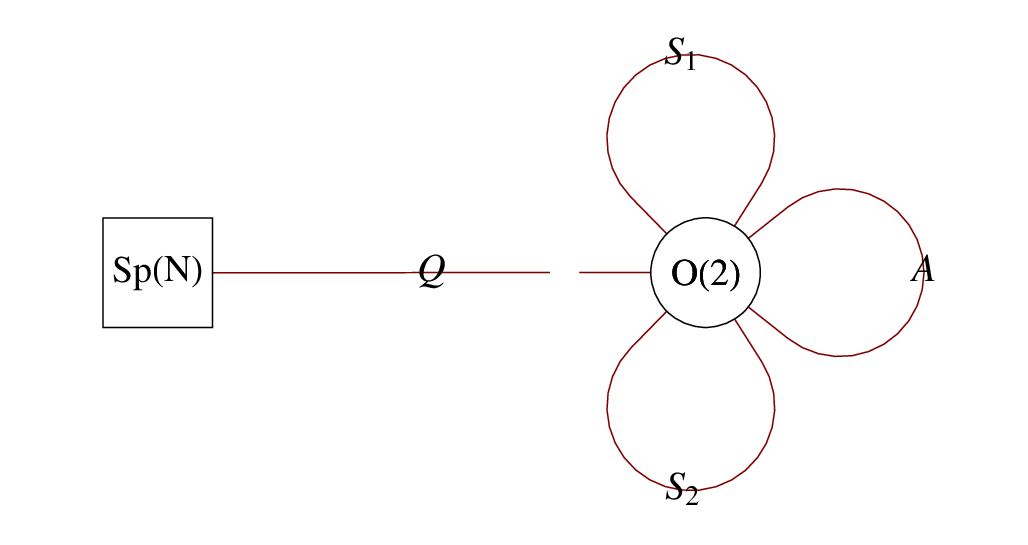}
\caption{The quiver diagram for the theory described by \fref{fig:N22SU2inst}, written in $\cN=1$ notation. The superpotential (setting mass terms to zero) is $W = J_{ij} Q_{~a}^i A_{ab}  Q_{~b}^{j} + \epsilon^{\alpha \beta}  (S_\alpha)_{ab} A_{bc} (S_\beta)_{ca}$.}
\label{fig:N12Sp1inst}
\end{center}
\end{figure}

The translation of the quiver diagram in \fref{fig:N22Sp1inst} to $\CN=1$ notation is depicted in \fref{fig:N12Sp1inst}.  The $O(2)$ adjoint (rank-2 antisymmetric) field $A$ comes from the scalar in the $\cN=2$ vector multiplet, and the chiral fields $S_1, S_2, Q$ comes from the $\cN=2$ hypermultiplets. 

We have a global symmetry $U(2)_{\BC^2} \times Sp(N)$, where $U(2)_{\BC^2} = U(1)_{\BC^2} \times SU(2)_{\BC^2}$ corresponds to the isometry of $\BC^2$ parametrised by the overall position of the instantons, and the second $Sp(N)$ corresponds to the square node in the quiver diagram.  The $O(2)$ rank-2 symmetric fields $S_\alpha = (S_1, S_2)$ transform as a doublet of $SU(2)_{\BC^2}$ and carry charge $+1$ under $U(1)_{\BC^2}$.  The fields $Q^i_{~a}$ (with $a = 1,2$ and $i=1,\ldots,2N$) transform under the bi-fundamental representation of $O(2) \times Sp(N)$; they are singlet under $SU(2)_{\BC^2}$ and carry charge $+1$ under $U(1)_{\BC^2}$.

Here and in the rest of the discussion, we use the indices $a,b,c=1,2$ for the gauge symmetry $O(2)$, $i, j, k=1, \ldots, 2N$ for the global symmetry $Sp(N)$, and $\alpha, \beta=1,2$ for the global symmetry $SU(2)_{\BC^2}$.  Due to $\CN=2$ supersymmetry, the superpotential is fixed to be (for simplicity, we set the mass terms to zero):
\bea \label{sup2sp1}
W = \epsilon_{ij} Q_{~a}^i A_{ab}  Q_{~b}^{j} + \epsilon^{\alpha \beta}  (S_\alpha)_{ab} A_{bc} (S_\beta)_{ca}~,
\eea
where the gauge indices are contracted using Kronecker delta's.
On the Higgs branch, the vacuum expectation values of $A_{ab}$ is zero.  Therefore, the $F$-terms are
\bea
0 = \partial_{A_{ab}} W = \epsilon_{ij} Q^i_{~a} Q^j_{~b}  + [S_1, S_2]_{ab}~.
\eea
This matrix equation transforms under the adjoint representation of $O(2)$.  

\subsection{Computing the Hilbert series for two $Sp(N)$ instantons}
The instructive part of this computation is to properly count invariants under the gauge group $O(2)$, and not $SO(2)$.

Let us first discuss some background (see, \eg, \cite{DK} for a review).  The Hilbert series that counts invariants under a discrete group $G$ can be computed using the Molien formula
\bea
H(t) = \frac{1}{|G|} \sum_{g \in G} \frac{1}{\det_V(\BU - t \rho(g))}~, \label{discreteMolien}
\eea
where $\rho: G \rightarrow GL(V)$ is a representation of $G$.  

This formula can be generalised to a compact connected Lie group $G$.  
The summation is replaced by an integral over the Haar measure $\ud \mu_G$, where we choose the normalisation such that $\int \ud \mu_G =1$.
Furthermore, the expression $1/\det_V(\BU - t \rho(g))$ only depends on the conjugacy class of $g$. 
Since every element $g \in G$ is conjugate to an element in a maximal torus of $G$ and all maximal tori are conjugate to each other,
the expression $1/\det_V(\BU - t \rho(g))$ can be reduced to $1/\det_V(\BU - t \sigma)$, 
where $\sigma$ is the action of a maximal torus on the dual space $V^*$.
Therefore, the Molien formula becomes
\bea
H(t) = \int_G \ud \mu_G \frac{1}{\det_V(\BU - t \sigma)}~. \label{MolienWeyl}
\eea
If there is a basis on $V^*$ such that the action of the maximal torus is diagonal, then the inverse of the determinant can be rewritten in terms of plethystic exponential.  For example, let us consider the first equality of \eref{FF2SUN}: The first term in the $\PE$ in numerator corresponds to $\sigma = \diag(z_1,~z_2,~ 1/z_1,~1/z_2)$ whereas the second term corresponds to $\sigma = \diag(1,~1,~z_1/z_2, z_2/z_1)$, and similarly for the denominator.    This is also the known as the {\bf Molien--Weyl formula}. It has been used to compute Hilbert series in various supersymmetric gauge theories.

Since $O(2)$ is compact but not connected, we need to further generalise \eref{discreteMolien} and \eref{MolienWeyl}.  There are two conjugacy classes $\CC_+$ and $\CC_-$, namely that containing the elements with determinant $+1$ (\ie~the subgroup $SO(2)$) and that containing the elements with determinant $-1$.  Note that $O(2)$ has a parity $\BZ_2 = O(2)/SO(2)$.  Hence, the Hilbert series can be written as
\bea
g_{2,Sp(N)} (t,x, \{ y_i \} ) = \frac{1}{2} \left[ H_+ (t,x,\{y_i \}) + H_- (t,x,\{y_i \}) \right]~, \label{MolienZ2}
\eea
where $H_+(t,x,\{y_i \})$ is the contribution from $\CC_+$ and $H_- (t,x,\{y_i \})$ is the contribution from $\CC_-$.
Here we take $t$ to be the fugacity of $U(1)_{\BC^2}$, $x$ to be the fugacity of $SU(2)_{\BC^2}$, $y_1, \ldots y_N$ to be the fugacity of $Sp(N)$, and $z$ to be the fugacity of $SO(2)$.

\paragraph{Computing $H_+$. }The Hilbert series $H_+$ can be computed as follows:
\bea \label{defHplus}
H_+ (t,x,y) = \frac{1}{2 \pi i} \oint_{|z| =1}  \frac{\ud z}{z} ~(1-t^2)  \PE \left[ t [1]_x \left( z^2 + 1 + \frac{1}{z^2} \right) + t [1,0, \ldots, 0]_y \left( z+ \frac{1}{z} \right) \right]~,
\eea
where the first term in the $\PE$ comes from the action 
\bea
\sigma^+_S = \diag( z^2,~1,~1/z^2)
\eea
of $\CC_+$ on the representation of the adjoint fields $S$, the second term in the $\PE$ comes from the action 
\bea
\sigma^+_Q =\diag(z,~1/z)
\eea
of $\CC_+$ on the representation of the quarks $Q$, and the factor $(1-t^2)$ comes from the $F$-term which transform as a singlet under $SO(2)$.
The character of the fundamental representation $[1,0,\ldots,0]$ of $Sp(N)$ can be taken as
\bea
[1,0,\ldots,0]_y = y_1 +y_1^{-1} + \sum_{i=1}^{N-1} \left( \frac{y_i}{y_{i+1}}+ \frac{y_{i+1}}{y_{i}} \right)~.
\eea
The result from \eref{defHplus} can be written as
\bea
H_+(t,x,y) = \frac{1}{(1-t x)\left(1-\frac{t}{x} \right)} \widetilde{H}_+ (t,x,y)~, \label{Hplus}
\eea
where $\widetilde{H}_+ (t,x,y)$ has an interpretation of a Hilbert series of the irreducible component of the Higgs branch of \fref{fig:N12Sp1inst} with $O(2)$ being replaced by $SO(2)$.

\paragraph{Computing $H_-$. } For the Hilbert series $H_-$, one has to be careful that element in $\CC_-$ do not commute with $\CC_+$, and hence actions of elements in $\CC_-$ on $V^*$ is not simultaneously diagonalisable with the actions $\sigma$'s in the previous paragraph.\footnote{We are indebted to Yuji Tachikawa for clarifying this issue and pointing this out to us.}  The parity $\BZ_2 = O(2)/SO(2)$ acts on the chiral fields as follows:
\bea
(Q^i_{~1}, Q^i_{~2}) \quad &\longrightarrow \quad (Q^i_{~2}, Q^i_{~1})~,  \nn \\
A_{12} \quad &\longrightarrow \quad -A_{12}~, \nn\\
S_{ab}  \quad &\longrightarrow \quad  S_{ab}~.
\eea
Thus, the action of $\CC_-$ on the representation of the quarks $Q$ is 
\bea
\sigma^-_Q = \begin{pmatrix} 0 & z \\ \frac{1}{z} & 0 \end{pmatrix}~,
\eea
the action of $\CC_-$ on the representation of the antisymmetric field $A$ is
\bea
\sigma^-_A = -1~,
\eea
and the action of $\CC_-$ on the representation of the symmetric field $S$ is
\bea
\sigma^-_S =  \left(
\begin{array}{ccc}
 0 & 0 & z^2 \\
 0 & 1 & 0 \\
 \frac{1}{z^2} & 0 & 0
\end{array}
\right)~.
\eea
Thus, we have
\bea
& H_- (t,x,y_1, \ldots, y_{N}) \nn \\
&= \frac{1}{2 \pi i}\oint_{|z|=1} \frac{\ud z}{z}~\frac{ \det(\BU - t^2 \sigma^-_A) }{ \left[ \det(\BU - t y_1^\pm \sigma^-_Q) \prod_{i=1}^{N-1} \det \left(\BU - t \left( y_i/ y_{i+1} \right)^{\pm}\sigma^-_Q \right)  \right] \det(\BU - t x \sigma^-_S) \det(\BU - \frac{t}{x} \sigma^-_S) } \nn \\
&= \frac{1}{2 \pi i} \oint_{|z|=1} \frac{\ud z}{z}~ \frac{1+t^2}{\left[(1-t^2 y_1^{\pm 2}) \prod_{i=1}^{N-1} (1-t^2\left( y_i/ y_{i+1} \right)^{\pm2})\right] (1-t x^\pm)(1-t^2 x^{\pm 2})} \nn \\
&=\frac{1+t^2}{\left[(1-t^2 y_1^{\pm 2}) \prod_{i=1}^{N-1} (1-t^2\left( y_i/ y_{i+1} \right)^{\pm2})\right] (1-t x^\pm)(1-t^2 x^{\pm 2})} \nn \\
&=: \frac{1}{1-t x^\pm} \widetilde{H}_- (t,x,y_1, \ldots, y_N)~, \label{Hminus}
\eea
where the shorthand notation $(1-t y^{\pm})$ stands for $(1-t y)(1-t y^{-1})$ and 
\bea
 \widetilde{H}_- (t,x,y_1, \ldots, y_N) = \frac{1+t^2}{\left[(1-t^2 y_1^{\pm 2}) \prod_{i=1}^{N-1} (1-t^2\left( y_i/ y_{i+1} \right)^{\pm2})\right] (1-t^2 x^{\pm 2})}~.
\eea
Observe that although $\sigma^-$ depend on the gauge fugacity $z$, the integrand in \eref{Hminus} does not.

\paragraph{The Hilbert series.} Using \eref{MolienZ2}, we obtain the Hilbert series for two $Sp(N)$ instantons, as required.  The Hilbert series of the reduced instanton moduli space is 
\bea \label{HS2SpN}
\widetilde{g}_{2, Sp(N)}(t,x,\{ y_i \}) = (1-tx^\pm) g_{2,Sp(N)}(t,x,\{ y_i \}) = \frac{1}{2} \left[ \widetilde{H}_+ (t,x,\{ y_i \})  +\widetilde{H}_- (t,x,\{ y_i \}) \right]~.
\eea
The generators at order $t^2$ transform under the $SU(2) \times Sp(N)$ representation $[2;0,\ldots,0]+[0;2,0,\ldots,0]=[\Adj_{SU(2)}; \mathbf{singlet}_{Sp(N)}]+ [\mathbf{singlet}_{SU(2)}; \Adj_{Sp(N)}]$ and those at order $t^3$ transform under the representation $[1;2,0,\ldots,0]= [\mathbf{\fun}_{SU(2)}; \Adj_{Sp(N)}]$.

\subsubsection{Example: Two $Sp(1)$ instantons}
In this section, we compute the Hilbert series of the moduli space of two $Sp(1)$ instantons on $\BR^4$.  Since $Sp(1)$ is isomorphic to $SU(2)$, one should expect that such a Hilbert series is equal to that of 2 $SU(2)$ instantons  \eref{includeCoM2SU2} and \eref{Irr2SU2}.  Let us present certain results here.

From \eref{defHplus}, we have 
\bea  \label{HplusSp2}
\widetilde{H}_+ (t,x,y) &=  \frac{1}{1-t^2} \sum_{m_2 = 0}^\infty   \sum_{n_2 = 0}^\infty  \sum_{n_3 = 0}^\infty  \Big \{ [2m_2+n_3; 2n_2+2n_3]_{x;y} t^{2m_2+2n_2+3n_3} \nn \\
& \hspace{4cm} +[2m_2+n_3+1;~2n_2+2n_3+2]_{x;y} t^{2m_2+2n_2+3n_3+3} \Big \}~.
\eea
From \eref{Hminus}, we have 
\bea  \label{HminusSp2}
\widetilde{H}_- (t,x,y) &= \frac{1+t^2}{(1-t^2 y^2)(1-t^2 y^{-2})(1-t^2 x^2)(1-t^2 x^{-2})}~.
\eea

Substituting \eref{HplusSp2} and \eref{HminusSp2} into \eref{MolienZ2}, we find that the Hilbert series $g_{2,Sp(1)}$ of 2 $Sp(1)$ instantons is equal to the Hilbert series $g_{2,SU(2)}$ of 2 $SU(2)$ instantons given by \eref{includeCoM2SU2}.  Explicitly, the Hilbert series of the reduced instanton moduli space is 
\bea
 \widetilde{g}_{2,Sp(1)} (t,x,y)  
 &= \frac{1}{2} \left[ \widetilde{H}_+ (t,x,y) + \widetilde{H}_- (t,x,y) \right] \nn \\
 &=  \frac{1}{1-t^4} \sum_{m_2 =0}^\infty  \sum_{n_2 =0}^\infty  \sum_{n_3 =0}^\infty  \Big \{ [2m_2+n_3; 2n_2+2n_3]_{x;y}~t^{2m_2+2n_2+3n_3}  \nn \\
& \quad +[2m_2+n_3+1; 2n_2+2n_3+2]_{x;y}~t^{2m_2+2n_2+3n_3+5} \Big \}~, \label{Irr2Sp1}
\eea

\subsection{General formula}
For higher $Sp(N)$, we can compute the Hilbert series case by case from \eref{HS2SpN}.  However, an expression obtained from the case by case computation does not lead to a clear generalisation for higher $N$ and other groups.  As for the case of $SU(N)$, we write the Hilbert series in terms of a character expansion.  This leads to a conjecture for the Hilbert series of the reduced 2 $Sp(N)$ instanton moduli space:
\bea
\widetilde{g}_{2,Sp(N)} (t,x,y_1, \ldots, y_N) &=  f(0; 0, \ldots, 0) +f(0;0,1,0,\ldots,0)t^{4} \nn \\
& \quad + \left[ f(1;2,0,0,\ldots,0) +  f(1;2,1,0,\ldots,0) \right] t^{5}~. \label{gen2SpN}
\eea
where the function $f$ is defined as
\bea
 f(a;b_1,b_2, \ldots, b_{N}) &=  \frac{1}{1-t^4} \sum_{m_2 =0}^\infty \sum_{n_2 =0}^\infty\sum_{n_3 =0}^\infty \sum_{n_4 =0}^\infty t^{2m_2+2n_2+3n_3+4n_4} \times \nn \\
& [2m_2+n_3+a; 2n_2+2n_3 +b_1,2n_4+b_2,b_3,...,b_N]
\eea
%\bea
%& \widetilde{g}_{2,Sp(N)} (t,x,y_1, \ldots, y_N) = \frac{1}{1-t^4} \sum_{m_2 =0}^\infty \sum_{n_2 =0}^\infty\sum_{n_3 =0}^\infty \sum_{n_4 =0}^\infty \nn \\
%& \Bigg\{ [2m_2+n_3; 2n_2+2n_3,2n_4,0,...,0]t^{2m_2+2n_2+3n_3+4n_4} \nn \\
%& +[2m_2+n_3; 2n_2+2n_3,2n_4+1,0,...,0] t^{2m_2+2n_2+3n_3+4n_4+4} \nn \\
%&+\Big([2m_2+n_3+1; 2n_2+2n_3+2,2n_4,0,...,0] \nn \\
%& +[2m_2+n_3+1; 2n_2+2n_3+2,2n_4+1,0,...,0] \Big) t^{2m_2+2n_2+3n_3+4n_4+5} \Bigg \}~.
%\eea

Setting $x = y_1= \ldots =y_N=1$ and performing the summations, we obtain the unrefined Hilbert series; let us present the results for a few values of $N$ below:
\bea
\widetilde{g}_{2,Sp(1)}(t) &=  \frac{1+t+3 t^2+6 t^3+8 t^4+6 t^5+8 t^6+6 t^7+3 t^8+t^9+t^{10}}{(1-t)^6 (1+t)^4 \left(1+t+t^2\right)^3}~, \nn \\
 \widetilde{g}_{2,Sp(2)}(t) &= \frac{1}{(1-t)^{10} (1+t)^6 \left(1+t+t^2\right)^5}\Big(1+t+8 t^2+23 t^3+50 t^4+95 t^5+177 t^6+222 t^7 \nn \\
& \quad +236 t^8+222 t^9+177 t^{10}+95 t^{11}+50 t^{12}+23 t^{13}+8 t^{14}+t^{15}+t^{16} \Big)~,\nn \\
 \widetilde{g}_{2,Sp(3)}(t) &= \frac{1}{(1-t)^{14} (1+t)^{10} (1+t+t^2)^7} \Big(1+3 t+20 t^2+87 t^3+308 t^4+921 t^5+2402 t^6+5115 t^7 \nn \\
 & \quad +9263 t^8+14650 t^9+20345 t^{10}+24503 t^{11}+26006 t^{12}+24503 t^{13}+20345 t^{14}+14650 t^{15} \nn \\
 & \quad +9263 t^{16}+5115 t^{17}+2402 t^{18}+921 t^{19}+308 t^{20}+87 t^{21}+20 t^{22}+3 t^{23}+t^{24} \Big)~.
\eea
Observe that the numerators of these Hilbert series are palindromic and the order of the poles at $t=1$ is equal to the complex dimension $4(N+1)-2 = 4N+2$ of the reduced instanton moduli space.  We conjecture that the unrefined Hilbert series for $N \geq 2$ takes the following form:
\bea
\widetilde{g}_{2,Sp(N)}(t) &= \frac{P_{8N}(t)}{(1-t)^{4N+2}(1+t)^{4N-2}(1+t+t^2)^{2N+1}}~,
\eea
where $P_{8N}(t)$ is a palindromic polynomial of order $8N$.

We use \eref{limxtgen} to test conjecture \eref{gen2SpN} for $N=1, \ldots, 4$ and the results are as required.

\subsubsection{The lattice structure}
\begin{figure}[htbp]
\begin{center}
\includegraphics[scale=1]{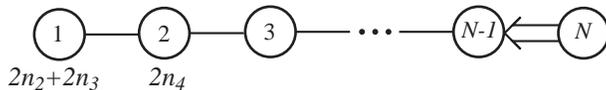}
\caption{The Dynkin diagram of $Sp(N)$. The labels in the first line indicate the representations corresponding to each node of the Dynkin label. The labels  below the nodes indicate the indices in the universal lattice $f(0;0,\ldots,0)$.  Observe that the lattices in \eref{gen2SpN} occupy only the first 2 nodes on the left of the Dynkin diagram.}
\label{DynkinCn}
\end{center}
\end{figure}

The universal lattice in $f(0;0,\ldots, 0)$ is generated by the highest weight vectors $[2;0,0, \ldots,0]+[0;2,0,\ldots,0]$ at order $t^2$, $[1;2,0,0,\ldots,0]$ at order $t^3$, and $[0;0, \ldots,0] + [0;0,2,0,\ldots,0]$ at order $t^4$. Note that the generators of the lattice at the former two orders are also the generators of the moduli space of two $Sp(N)$ instantons.  The representations corresponding to each node of the Dynkin diagram and the indices corresponding to the universal lattice are depicted in \fref{DynkinCn}.  Observe that there is no other lattices than the universal one, and that only the first two nodes on the left of the Dynkin label are occupied.

%\paragraph{Dimension of the lattice and dimension of the moduli space.} We can estimate the dimension of the moduli space from the dimension of the lattices and vice-versa.  Note that 
%\bea
%\dim [n, n, 0, \ldots, 0]_{Sp(N)} \sim n^{4{N-4}}~, \qquad n \rightarrow \infty,~N \geq 2~.
%\eea
%Then, as $t \rightarrow 1$, we estimate the most singular part of the unrefined Hilbert series from \eref{gen2SpN} as follows:
%\bea
%& \frac{1}{1-t^4} \left[  \sum_{r=0}^\infty \dim[r]_{SU(2)} t^r \right] \times \left[  \sum_{a_1 =0}^\infty {a_1}^{4N-4} t^{a_1} \sum_{a_2,a_3,a_4 =0}^\infty t^{a_2+ a_3+ a_4} \right] \nn \\
%& \sim \frac{1}{1-t} \times \frac{1}{(1-t)^2}  \times \frac{1}{(1-t)^{4N-3}} \times  \frac{1}{(1-t)^3}  \nn \\
%& = \frac{1}{(1-t)^{4N+3}}~.
%\eea
%Observe that the order of the pole at $t=1$ is $4N+2$.  This agrees with the complex dimension of the reduced 2 $Sp(N)$ instantons. 

\section{Two $SO(N)$ instantons} \label{sec:2SONinsts}
The ADHM data are given by a 4d $\CN=2$ gauge theory whose quiver diagram is depicted in \fref{fig:N22SU2inst}.  We focus on the Higgs branch of this theory.  The moment map equations, which consists of $F$ and $D$ terms in 4d $\CN=1$ language, give rise to a hyperK\"ahler quotient of the moduli space of two $SO(N)$ instantons on $\BR^4$.

\begin{figure}[htbp]
\begin{center}
\includegraphics[height=0.8 in]{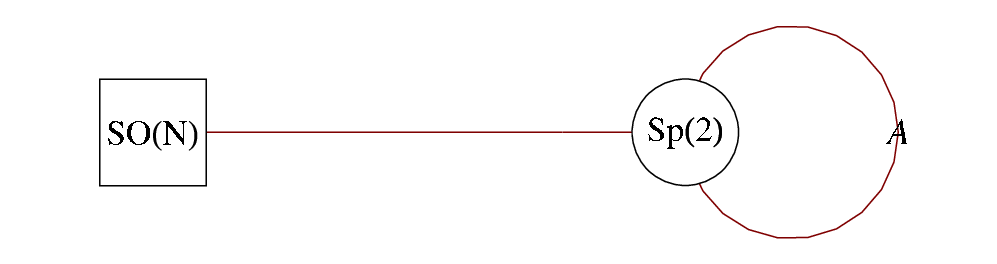}
\caption{The quiver diagram of a 4d $\CN=2$ gauge theory with the gauge group $Sp(2)$ and a global symmetry $SO(N)$. The matter content consists of a bifundamental hypermultiplet of $Sp(2) \times SO(N)$, and a rank-2 antisymmetric hypermultiplet of the $Sp(2)$ gauge group. }
\label{fig:N22SONinst}
\end{center}
\end{figure}

\begin{figure}[htbp]
\begin{center}
\includegraphics[scale=0.7]{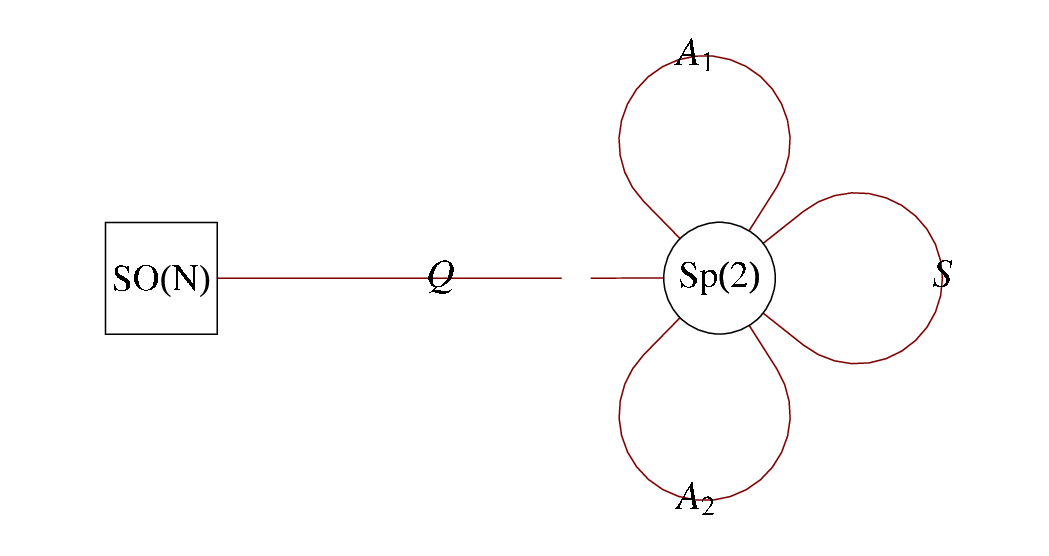}
\caption{The quiver diagram for the theory described by \fref{fig:N22SU2inst}, written in $\cN=1$ notation. The superpotential (setting mass terms to zero) is $W = \delta_{ij} Q_{~a}^i S_{ab}  Q_{~b}^{j} + \epsilon^{\alpha \beta}  (A_\alpha)_{ab} S_{bc} (A_\beta)_{ca}$.}
\label{fig:N12SONinst}
\end{center}
\end{figure}

The translation of the quiver diagram in \fref{fig:N22SONinst} to $\CN=1$ notation is depicted in \fref{fig:N12SONinst}.  The $Sp(2)$ adjoint (rank-2 symmetric) field $S$ comes from the scalar in the $\cN=2$ vector multiplet, and the chiral fields $A_1, A_2, Q$ comes from the $\cN=2$ hypermultiplets. 

We have a global symmetry $U(2)_{\BC^2} \times SO(N)$, where $U(2)_{\BC^2} = U(1)_{\BC^2} \times SU(2)_{\BC^2}$ corresponds to the isometry of $\BC^2$ parametrised by the overall position of the instantons, and the second $SO(N)$ corresponds to the square node in the quiver diagram.  The $Sp(2)$ rank-2 antisymmetric fields $A_\alpha = (A_1, A_2)$ transform as a doublet of $SU(2)_{\BC^2}$ and carry charge $+1$ under $U(1)_{\BC^2}$.  The fields $Q^i_{~a}$ (with $a = 1,2,3,4$ and $i=1,\ldots,N$) transform under the bi-fundamental representation of $Sp(2) \times SO(N)$; they are singlet under $SU(2)_{\BC^2}$ and carry charge $+1$ under $U(1)_{\BC^2}$.

Here and in the rest of the discussion, we use the indices $a,b,c=1,\ldots,4$ for the gauge symmetry $Sp(2)$, $i, j, k=1, \ldots, N$ for the global symmetry $SO(N)$, and $\alpha, \beta=1,2$ for the global symmetry $SU(2)_{\BC^2}$.  Due to $\CN=2$ supersymmetry, the superpotential is fixed to be (for simplicity, we set the mass terms to zero):
\bea \label{sup2soN}
W = \delta_{ij} Q_{~a}^i S_{ab}  Q_{~b}^{j} + \epsilon^{\alpha \beta}  (A_\alpha)_{ab} S_{bc} (A_\beta)_{ca}~,
\eea
where the gauge indices are contracted using Kronecker delta's.
On the Higgs branch, the vacuum expectation values of $A_{ab}$ is zero.  Therefore, the $F$-terms are
\bea
0 = \partial_{S_{ab}} W = Q^i_{~a} Q^i_{~b}  + [A_1, A_2]_{ab}~.
\eea
This matrix equation transforms under the adjoint representation of $Sp(2)$.  

Therefore, we can write down the Hilbert series of the space of the $F$-term solutions as 
\bea
g^{\ff} (t,x,y_1, \ldots, y_N;z_1,z_2) &= \frac{\PE\left[ [1,0]_z [1,0, \ldots,0]_y t +  ([0,1]_z+1) [1]_x t \right]}{\PE \left[ [2,0]_{z} t^2   \right]}~,
%&= \frac{\prod_{1 \leq a, b \leq 2}  \left(1- \frac{z_a}{z_b} t^2 \right) }{ \left[ \prod_{i = 0}^N \prod_{a=1}^2  \left(1- t z_a \frac{y_i}{y_{i+1}}  \right) \left(1- t z_a ^{-1}\frac{y_{i+1}}{y_{i}}  \right)  \right] \left[ \prod_{\delta = \pm 1} \prod_{1 \leq a, b \leq 2}  \left(1- t \frac{z_a}{z_b} x^\delta  \right)  \right]} \nn \\
%& \qquad \text{(with $y_0=y_{N+1} =1$)}~, 
\eea
where $t$ is a fugacity of $U(1)_{\BC^2}$, $x$ is a fugacity of $SU(2)_{\BC^2}$ and $y_1, \ldots, y_{N-1}$ are fugacities of $SU(N)$.

The Hilbert series of the Higgs branch of the quiver gauge theory depicted in \fref{fig:N22SU2inst} is given by.
\bea
g_{2,SO(N)}(t,x,y_1, \ldots, y_N) &= \int \ud \mu_{Sp(2)} (z_1,z_2)~ g^{\ff} (t,x,y_1,\ldots,y_N; z_1,z_2)~.
%&= \frac{1}{2} \oint_{|z_1| =1} \frac{\ud z_1}{z_1} \oint_{|z_2| =1} \frac{\ud z_1}{z_2} \left(\frac{1}{z_1}-\frac{1}{z_2} \right)(z_1-z_2) \times \nn \\
%& \quad \frac{\prod_{1 \leq a, b \leq 2}  \left(1- \frac{z_a}{z_b} t^2 \right) }{ \left[ \prod_{i = 0}^N \prod_{a=1}^2  \left(1- t z_a \frac{y_i}{y_{i+1}}  \right) \left(1- t z_a ^{-1}\frac{y_{i+1}}{y_{i}}  \right)  \right] \left[ \prod_{\delta = \pm 1} \prod_{1 \leq a, b \leq 2}  \left(1- t \frac{z_a}{z_b} x^\delta  \right)  \right]} \nn \\
%& \quad \text{(with $y_0=y_{N+1} =1$)}~,
\label{HS2SONint}
\eea
where the Haar measure of $U(2)$ is given by
\bea \label{HaarSp2}
\int \ud \mu_{Sp(2)} (z_1,z_2) = \frac{1}{(2 \pi i)^2} \oint_{|z_1| =1} \frac{\ud z_1}{z_1} \oint_{|z_2| =1} \frac{\ud z_1}{z_2} (1-z_1^2)(1-z_2)(1-z_1^2 z_2^{-1})(1- z_2^2 z_1^{-2}) ~.
\eea

\subsection{General formula}
For higher $SO(N)$, we can compute the Hilbert series case by case from \eref{HS2SONint}.  However, an expression obtained from the case by case computation does not lead to a clear generalisation for higher $N$ and other groups.  As for the case of $SU(N)$, we write the Hilbert series in terms of a character expansion. This leads to a conjecture for the Hilbert series of the reduced 2 $SO(N)$ instanton moduli space:
\bea  \label{gen2SON}
& \widetilde{g}_{2,SO(N)} (t,x,y_1, \ldots, y_r) =  \sum_{k_8=0}^\infty \Bigg 
\{ f(0;0,2k_8,0,\ldots, 0) t^{8k_8} + f(1;0,2k_8+1,0,\ldots, 0) t^{8k_8+5}  \nn \\
& \quad + f(1;1,2k_8,0,\ldots, 0) t^{8k_8+7}+f(0;1,2k_8+1,1,0, \ldots,0) t^{8 k_8+10} \Bigg \}  \nn \\
& \quad  + \sum_{k_5=0}^\infty \Bigg \{ f(k_5+1; k_5+1, 0, k_5+1, 0, \ldots, 0) t^{5 k_5+5} +f(k_5+2; k_5+2, 0, k_5+2, 0, \ldots, 0) t^{5 k_5+12} \Bigg \} ~.
%& + [2 m_2+n_3+1;2 n_4+n_6,n_2+n_3+2 k_8+1,n_6+2 n_8,m_4,0,\ldots,0] t^{2 m_2+2 n_2+3 n_3+4 n_4+4 m_4+6 n_6+8 n_8+8 k_8+5} \nn \\
%& +[2 m_2+n_3+1;2 n_4+n_6+1,n_2+n_3+2 k_8,n_6+2 n_8+1,m_4,0,\ldots,0] t^{2 m_2+2 n_2+3 n_3+4 n_4+4 m_4+6 n_6+8 n_8+8 k_8+7} \nn \\
%& +[2 m_2+n_3;2 n_4+n_6+1,n_2+n_3+2 k_8+1,n_6+2 n_8+1,m_4,0,\ldots,0] t^{2 m_2+2 n_2+3 n_3+4 n_4+4 m_4+6 n_6+8 n_8+8 k_8+10 } \Bigg\} \nn \\
%& + \frac{1}{1-t^4} \sum_{m_2 =0}^\infty \sum_{n_2 =0}^\infty\sum_{n_3 =0}^\infty \sum_{n_4 =0}^\infty\sum_{m_4 =0}^\infty \sum_{k_5 =0}^\infty  \sum_{n_6 =0}^\infty  \sum_{n_8 =0}^\infty \nn \\
%& \Bigg \{ [2 m_2+n_3+k_5+1;2 n_4+n_6+k_5+1,n_2+n_3,n_6+2 n_8+k_5+1,m_4,0, \ldots,0] t^{2 m_2+2 n_2+3 n_3+4 n_4+4 m_4+6 n_6+5 k_5+8 n_8+5} \nn \\
%& +[2 m_2+n_3+k_5+2;2 n_4+n_6+k_5+2,n_2+n_3,n_6+2 n_8+k_5+2,m_4,0, \ldots,0] t^{2 m_2+2 n_2+3 n_3+4 n_4+4 m_4+6 n_6+8 n_8+5 k_5+12} \Bigg \}~,
\eea
where $r$ is the rank of $SO(N)$ and the function $f$ is defined as
\bea
&f(a;b_1, \ldots, b_r) = \frac{1}{1-t^4} \sum_{m_2 =0}^\infty \sum_{n_2 =0}^\infty\sum_{n_3 =0}^\infty \sum_{n_4 =0}^\infty\sum_{m_4 =0}^\infty \sum_{n_6 =0}^\infty  \sum_{n_8 =0}^\infty  \nn\\
& [2 m_2 + n_3 + a; 2 n_4 + n_6 +b_1, n_2 + n_3+b_2 , n_6 + 2 n_8+b_3, m_4+b_4, b_5, \ldots, b_{r}] t^{2 m_2 + 2 n_2 + 3 n_3 + 4 n_4 + 4 m_4 + 6 n_6 + 8 n_8 }~.
\eea
Note that there are two sets of generators of the moduli space: Those at order $t^2$ transform under the $SU(2) \times SO(N)$ representation $[\Adj_{SU(2)}; \mathbf{singlet}_{SO(N)}]+ [\mathbf{singlet}_{SU(2)}; \Adj_{SO(N)}]$ and those at order $t^3$ transform under the representation $[\mathbf{\fun}_{SU(2)}; \Adj_{SO(N)}]$.

We conjecture that for $SO(N)$, with $N \geq 10$, the unrefined Hilbert series is given by
\bea
\widetilde{g}_{2,SO(N)} (t,x=1,\{y_i =1\}) = \frac{ P_{8N-30}(t) }{(1-t)^{4N-10}(1+t)^{4N-20}(1+t+t^2)^{2N-5}}~,
\eea
where $P_{8N-30}(t)$ is a palindromic polynomial of degree $8N-30$.

\subsubsection{The lattice structure}
\begin{figure}[htbp]
\begin{center}
\includegraphics[scale=0.8]{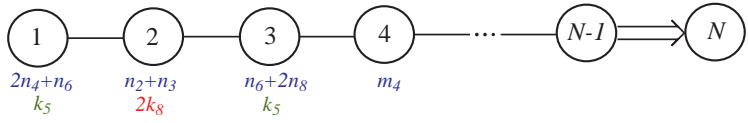}
\caption{The Dynkin diagram of $SO(2N+1)$.  The labels in black indicate ordering of the nodes; the one with number $n$ can be associated with the representation $[0,\ldots, 0, 1, 0, \ldots, 0]$ of $SO(2N+1)$, with $1$ in the $n$-th position from the left. The labels in blue indicate the indices in the universal lattice $f(0;0,\ldots,0)$. The labels in red and green indicate the indices in the non-universal lattices.}
\label{DynkinBn}
\end{center}
\end{figure}

\begin{figure}[htbp]
\begin{center}
\includegraphics[scale=0.8]{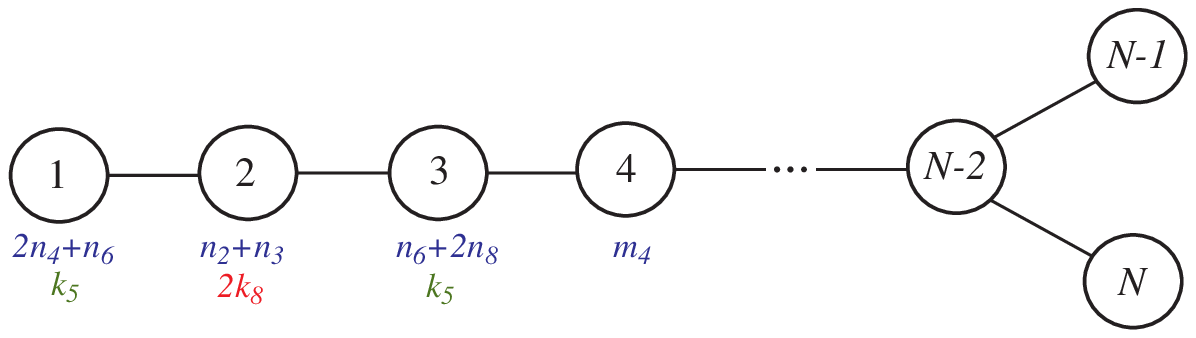}
\caption{The Dynkin diagram of $SO(2N)$.  The labels in black indicate ordering of the nodes; the one with number $n$ can be associated with the representation $[0,\ldots, 0, 1, 0, \ldots, 0]$ of $SO(2N)$, with $1$ in the $n$-th position from the left. The labels in blue indicate the indices in the universal lattice $f(0;0,\ldots,0)$. The labels in red and green indicate the indices in the non-universal lattices.}
\label{DynkinDn}
\end{center}
\end{figure}

General formula \eref{gen2SON} consists of three building blocks:
\bi
\item  {\bf The generators of the universal lattice.}  This is contained in $f(0;0,\ldots, 0)$.  They are associated with the indices $n$'s and $m$'s, whose subscripts indicate the order of $t$ at which each of them appears.  We tabulate them in \tref{genuniv2SON} and depict them in Figures \ref{DynkinBn}, \ref{DynkinDn}.
\item  {\bf The generators of the non-universal lattice.} They appear at orders 5 and 8 and are indicated by the indices $k_5$ and $k_8$.  We tabulate these in \tref{gennonuniv2SON} and depict them in Figures \ref{DynkinBn}, \ref{DynkinDn}.
\item  {\bf The shifts} from the universal and non-universal lattices at various orders of $t$.  These are summarised in \tref{shifts2SON}.
\ei

\begin{table}[htdp]
\footnotesize
\begin{center}
\begin{tabular}{|c||c|c|}
\hline
Order of $t$ & Dynkin labels for generic $SO(N)$ & Name of representation \\
\hline
2 & $[2;0, \ldots, 0]$ and $[0;0,1,0, \ldots,0]$ & $[2; \mathbf{singlet}_{SO(N)}]$ and $\mathbf{Adj}_{SO(N)}$ \\
\hline
3 & $[1;0,1,0, \ldots,0]$ & $[1; \mathbf{Adj}_{SO(N)}]$ \\
\hline
4 & $[0;2,0, \ldots, 0]$, $[0;0,0,0,1,0, \ldots, 0]$ and $[0;0, \ldots, 0]$ &  $\Sym^2 \mathbf{fund}_{SO(N)}$, $\wedge^4 \mathbf{fund}_{SO(N)}$ and $\mathbf{singlet}$ \\
\hline
6 & $[0;1,0,1,0, \ldots,0]$ & HWR in $\fun_{SO(N)} \otimes \wedge^3 \fun_{SO(N)}$ \\
\hline
8 & $ [0;0,0,2,0,\ldots,0]$ & HWR in $\Sym^2  \wedge^3 \fun_{SO(N)} $ \\
\hline
\end{tabular}
\end{center}
\caption{Generators of the universal lattice for the Hilbert series of 2 $SO(N)$ instantons. The acronym `HWR' stands for the highest weight representation; for example,  `HWR in $\fun_{SO(N)} \otimes \wedge^3 \fun_{SO(N)}$' means the highest weight representation that appears in the tensor product decomposition of $\fun_{SO(N)} \otimes \wedge^3 \fun_{SO(N)}$.}
\label{genuniv2SON}
\end{table}%

\begin{table}[htdp]
\begin{center}
\footnotesize
\begin{tabular}{|c||c|c|}
\hline
Order of $t$ & Dynkin labels for generic $SO(N)$ & Name of representation \\
\hline
5 & $[1;1,0,1,0,\ldots,0] $ & $[1; \text{HWR in}~\fun_{SO(N)} \otimes \wedge^3 \fun_{SO(N)}]$ \\
\hline
8 & $[0;0,2,0,\ldots,0] $ & HWR in $\Sym^2 \Adj_{SO(N)}$ \\
\hline
\end{tabular}
\end{center}
\caption{Generators of the non-universal lattice for the Hilbert series of 2 $SO(N)$ instantons.}
\label{gennonuniv2SON}
\end{table}%

\begin{table}[htdp]
\begin{center}
\footnotesize
\begin{tabular}{|c||c|c|}
\hline
Order of $t$ & Shift & Name of the representation \\
\hline
0 & $[0; 0, \ldots, 0] $ & $[0; \mathbf{singlet}_{SO(N)}]$ \\
\hline
5 (universal lattice) & $[1;0,1,0, \ldots, 0]$ & $[1; \Adj_{SO(N)}]$\\
\hline
5 (non-universal lattice) & $[1;1,0,1,0, \ldots,0] $ & $[1; \text{HWR in}~\fun_{SO(N)} \otimes \wedge^3 \fun_{SO(N)} ]$ \\
\hline
7 & $[1;1,0,1,0, \ldots, 0] $ & $[1; \text{HWR in}~\fun_{SO(N)} \otimes \wedge^3 \fun_{SO(N)} ]$\\
\hline 
10 & $[0;1,1,1,0,\dots,0]$ & $[1; \text{HWR in}~ \fun_{SO(N)} \otimes \wedge^2 \fun_{SO(N)} \otimes \wedge^3 \fun_{SO(N)} ]$ \\
\hline
12 & $[2;2,0,2,0, \ldots,0]$ & $[1; \text{HWR in}~\Sym^2 \fun_{SO(N)} \otimes \Sym^2 (\wedge^3 \fun_{SO(N)}) ]$ \\
\hline
\end{tabular}
\end{center}
\caption{Shifts from the universal and non-universal lattices at various orders of $t$.}
\label{shifts2SON}
\end{table}%

\subsection{Special cases of low rank groups}
Note that general formula \eref{gen2SON} takes its exact form for $N \geq 11$.   For smaller $N$, the general formula receives some corrections due to irregularities of the highest weight representations in tensor product decompositions.  We discuss this in details below.

\subsubsection{The case of $ SO(10)$} 
\begin{figure}[htbp]
\begin{center}
\includegraphics[scale=0.8]{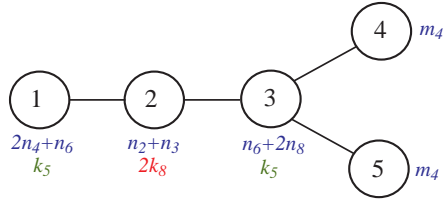}
\caption{The Dynkin diagram of $SO(10)$.  The labels in black indicate ordering of the nodes; the one with number $n$ can be associated with the representation $[0,\ldots, 0, 1, 0, \ldots, 0]$ of $SO(10)$, with $1$ in the $n$-th position from the left. The labels in blue indicate the indices in the universal lattice $f(0;0,\ldots,0)$. The labels in red and green indicate the indices in the non-universal lattices.}
\label{DynkinD5}
\end{center}
\end{figure}

When working with $SO(10)$, representations appearing in general formula \eref{gen2SON} receive some corrections. Let us first focus on the universal lattice.
\bea
\small
\wedge^3 \fun_{SO(10)} &= \wedge^3 [1,0,0,0,0] = [0,0,1,0,0]~, \nn \\
\wedge^4 \fun_{SO(10)} &= [0,0,0,1,1]~, \nn \\
\fun_{SO(10)} \otimes \wedge^3 \fun_{SO(10)} &= \mathbf{[1,0,1,0,0]} +[0,0,0,1,1] +[0,1,0,0,0]~, \nn\\
\Sym^2  \wedge^3 \fun_{SO(10)} &= \mathbf{[0,0,2,0,0]} + [1,0,0,2,0] +[1,0,0,0,2] +[0,2,0,0,0] +
[0,0,0,1,1] \nn \\
& \quad +[2,0,0,0,0] +[0,0,0,0,0]~, \nn \\
\Sym^2 \Adj_{SO(10)} &= \mathbf{[0,2,0,0,0]}+[0,0,0,1,1] +[2,0,0,0,0] +[0,0,0,0,0]~, \label{tensordecompSO10}
\eea
where the representation with the highest weight is denoted in boldface.  Comparing \eref{tensordecompSO10} with Tables \ref{genuniv2SON}, we see that the corrections to \eref{gen2SON}, when working with $SO(10)$, come from $\wedge^4 \fun_{SO(10)}$.  In particular, the corresponding generator is $[0,0,0,1,1]$, instead of $[0,0,0,0,1]$ as written in \eref{gen2SON}. Hence, the index $m_4$ appears in the fourth and fifth positions of the $SO(10)$ Dynkin label, as shown in \fref{DynkinD5}.

If one performs tensor product decompositions according to Tables \ref{gennonuniv2SON} and \ref{shifts2SON}, one finds that there are no corrections to the Dynkin labels appearing in the non-universal lattice and no corrections to the shifts.

The Hilbert series of the reduced 2 $SO(10)$ instanton moduli space is conjectured to be
{\footnotesize
\bea  \label{gen2SO10}
 \widetilde{g}_{2,SO(10)} (t,x,y_1, \ldots, y_5) &=  \sum_{k_8=0}^\infty \Bigg 
\{ f(0;0,2k_8,0,0, 0) t^{8k_8} + f(1;0,2k_8+1,0,0, 0) t^{8k_8+5} + f(1;1,2k_8,0,0, 0) t^{8k_8+7} \nn \\
& \quad +f(0;1,2k_8+1,1,0, 0,0) t^{8 k_8+10} \Bigg \} + \sum_{k_5=0}^\infty \Bigg \{ f(k_5+1; k_5+1, 0, k_5+1, 0, 0) t^{5 k_5+5}  \nn\\
& \quad +f(k_5+2; k_5+2, 0, k_5+2, 0, 0) t^{5 k_5+12} \Bigg \} ~.
\eea}
where the function $f$ is defined as
\bea
&f(a;b_1, \ldots, b_5) = \frac{1}{1-t^4} \sum_{m_2 =0}^\infty \sum_{n_2 =0}^\infty\sum_{n_3 =0}^\infty \sum_{n_4 =0}^\infty\sum_{m_4 =0}^\infty \sum_{n_6 =0}^\infty  \sum_{n_8 =0}^\infty t^{2 m_2 + 2 n_2 + 3 n_3 + 4 n_4 + 4 m_4 + 6 n_6 + 8 n_8 } \times \nn\\
& [2 m_2 + n_3 + a; 2 n_4 + n_6 +b_1, n_2 + n_3+b_2 , n_6 + 2 n_8+b_3, m_4+b_4, m_4+b_5] ~.
\eea

The unrefined Hilbert series is given by
\bea
 \widetilde{g}_{2,SO(10)} (t,x,y)  
 &= \frac{1}{(1-t)^{30} (1+t)^{20} \left(1+t+t^2\right)^{15}}\times \Big( 1+5 t+43 t^2+250 t^3+1270 t^4+5736 t^5+23112 t^6 \nn \\
 & +83328 t^7+273201 t^8+817951 t^9+2246830 t^{10}+5691011 t^{11}+13349064 t^{12}+29079663 t^{13} \nn \\
 & +59001462 t^{14}+111786869 t^{15}+198188012 t^{16}+329355491 t^{17}+513867799 t^{18} \nn \\
 & +753728586 t^{19}+1040437264 t^{20}+1352815918 t^{21}+1658110082 t^{22}+1916789664 t^{23} \nn \\
 & +2090643428 t^{24}+2151958688 t^{25}+2090643428 t^{26}+ \text{palindrome up to $t^{50}$ \Big)} ~.
\eea

\subsubsection{The case of $ SO(9)$} 
\begin{figure}[htbp]
\begin{center}
\includegraphics[scale=0.9]{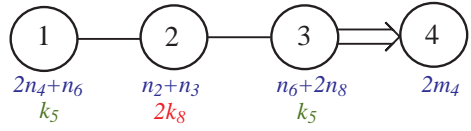}
\caption{The Dynkin diagram of $SO(9)$.  The labels in black indicate ordering of the nodes; the one with number $n$ can be associated with the representation $[0,\ldots, 0, 1, 0, \ldots, 0]$ of $SO(9)$, with $1$ in the $n$-th position from the left. The labels in blue indicate the indices in the universal lattice $f(0;0,\ldots,0)$. The labels in red and green indicate the indices in the non-universal lattices.}
\label{DynkinB4}
\end{center}
\end{figure}

When working with $SO(9)$, representations appearing in general formula \eref{gen2SON} receive some corrections. Let us first focus on the universal lattice. We focus on the following tensor product decompositions
\bea
\wedge^3 \fun_{SO(9)} &= \wedge^3 [1,0,0,0] = [0,0,1,0]~, \nn \\
\wedge^4 \fun_{SO(9)} &= [0,0,0,2]~, \nn \\
\fun_{SO(9)} \otimes \wedge^3 \fun_{SO(9)} &= \mathbf{[1,0,1,0]} +[0,0,0,2] +[0,1,0,0]~, \nn\\
\Sym^2  \wedge^3 \fun_{SO(9)} &= \mathbf{[0,0,2,0]} +[1,0,0,2] +[0,2,0,0] +[0,0,0,2] +[2,0,0,0] +[0,0,0,0]~, \nn \\
\Sym^2 \Adj_{SO(9)} &= \mathbf{[0,2,0,0]} +[0,0,0,2] +[2,0,0,0] +[0,0,0,0]~, \label{tensordecompSO9}
 \eea
where the representation with the highest weight is denoted in boldface.  Comparing \eref{tensordecompSO9} with Tables \ref{genuniv2SON}, we see that the corrections to \eref{gen2SON}, when working with $SO(9)$, come from $\wedge^4 \fun_{SO(9)}$.  In particular, the corresponding generator is $[0,0,0,2]$, instead of $[0,0,0,1]$ as written in \eref{gen2SON}. Hence, the index $m_4$ appears in the fourth position of the $SO(9)$ Dynkin label and with coefficient 2, as shown in \fref{DynkinB4}.

If one performs tensor product decompositions according to Tables \ref{gennonuniv2SON} and \ref{shifts2SON}, one finds that there are no corrections to the Dynkin labels appearing in the non-universal lattice and no corrections to the shifts.

Explicitly, the function $f$ for 2 $SO(9)$ instantons is given by
\bea
&f(a;b_1, \ldots, b_r) = \frac{1}{1-t^4} \sum_{m_2 =0}^\infty \sum_{n_2 =0}^\infty\sum_{n_3 =0}^\infty \sum_{n_4 =0}^\infty\sum_{m_4 =0}^\infty \sum_{n_6 =0}^\infty  \sum_{n_8 =0}^\infty   t^{2 m_2 + 2 n_2 + 3 n_3 + 4 n_4 + 4 m_4 + 6 n_6 + 8 n_8 } \times \nn\\
& \qquad  [2 m_2 + n_3 + a; 2 n_4 + n_6 +b_1, n_2 + n_3+b_2 , n_6 + 2 n_8+b_3, 2m_4+b_4]~,
\eea
and the Hilbert series of the reduced instanton moduli space can be written as
\bea  \label{gen2SO9}
 \widetilde{g}_{2,SO(9)} (t,x,y_1, \ldots, y_4) &=  \sum_{k_8=0}^\infty \Bigg 
\{ f(0;0,2k_8,0,0) t^{8k_8} + f(1;0,2k_8+1,0,0) t^{8k_8+5} + f(1;1,2k_8,0, 0) t^{8k_8+7} \nn \\
& \quad +f(0;1,2k_8+1,1,0) t^{8 k_8+10} \Bigg \} + \sum_{k_5=0}^\infty \Bigg \{ f(k_5+1; k_5+1, 0, k_5+1, 0) t^{5 k_5+5}  \nn\\
& \quad +f(k_5+2; k_5+2, 0, k_5+2, 0) t^{5 k_5+12} \Bigg \} ~.
\eea
%
%
%
%\bea \label{hs2SO9}
%& \widetilde{g}_{2,SO(9)} (t,x,y_1, \ldots, y_4) = \frac{1}{1-t^4} \sum_{m_2 =0}^\infty \sum_{n_2 =0}^\infty\sum_{n_3 =0}^\infty \sum_{n_4 =0}^\infty\sum_{m_4 =0}^\infty \sum_{n_6 =0}^\infty  \sum_{n_8 =0}^\infty  \sum_{k_8 =0}^\infty \nn\\
%& \Bigg\{ [2 m_2 + n_3; 2 n_4 + n_6, n_2 + n_3 + 2 k_8, n_6 + 2 n_8, 2m_4] t^{2 m_2 + 2 n_2 + 3 n_3 + 4 n_4 + 4 m_4 + 6 n_6 + 8 n_8 + 8 k_8} \nn \\
%& + [2 m_2+n_3+1;2 n_4+n_6,n_2+n_3+2 k_8+1,n_6+2 n_8,2m_4] t^{2 m_2+2 n_2+3 n_3+4 n_4+4 m_4+6 n_6+8 n_8+8 k_8+5} \nn \\
%& +[2 m_2+n_3+1;2 n_4+n_6+1,n_2+n_3+2 k_8,n_6+2 n_8+1,2m_4] t^{2 m_2+2 n_2+3 n_3+4 n_4+4 m_4+6 n_6+8 n_8+8 k_8+7} \nn \\
%& +[2 m_2+n_3;2 n_4+n_6+1,n_2+n_3+2 k_8+1,n_6+2 n_8+1,2m_4] t^{2 m_2+2 n_2+3 n_3+4 n_4+4 m_4+6 n_6+8 n_8+8 k_8+10 } \Bigg\} \nn \\
%& + \frac{1}{1-t^4} \sum_{m_2 =0}^\infty \sum_{n_2 =0}^\infty\sum_{n_3 =0}^\infty \sum_{n_4 =0}^\infty\sum_{m_4 =0}^\infty \sum_{k_5 =0}^\infty  \sum_{n_6 =0}^\infty  \sum_{n_8 =0}^\infty \nn \\
%& \Bigg \{ [2 m_2+n_3+k_5+1;2 n_4+n_6+k_5+1,n_2+n_3,n_6+2 n_8+k_5+1,2m_4] t^{2 m_2+2 n_2+3 n_3+4 n_4+4 m_4+6 n_6+5 k_5+8 n_8+5}\nn \\
%& +[2 m_2+n_3+k_5+2;2 n_4+n_6+k_5+2,n_2+n_3,n_6+2 n_8+k_5+2,2m_4] t^{2 m_2+2 n_2+3 n_3+4 n_4+4 m_4+6 n_6+8 n_8+5 k_5+12} \Bigg \}~.
%\eea
The unrefined Hilbert series is 
\bea
\widetilde{g}_{2,SO(9)} (t,x,y_1, \ldots, y_4) &= \frac{1}{(1-t)^{26} (1+t)^{16} \left(1+t+t^2\right)^{13}} \Big( 1+3 t+29 t^2+138 t^3+605 t^4+2373 t^5+8226 t^6 \nn \\
& +25281 t^7+70816 t^8+180223 t^9+418543 t^{10}+893096 t^{11}+1758304 t^{12}+3197816 t^{13} \nn \\
& +5390623 t^{14}+8445331 t^{15}+12313069 t^{16}+16725702 t^{17}+21203735 t^{18}+25107189 t^{19} \nn \\
& +27775791 t^{20}+28722684 t^{21}+27775791 t^{22} + \text{palindrome up to $t^{42}$} \Big)~.
\eea

\subsubsection{The case of $SO(8)$} 
Let us first focus on the universal lattice. In order to determine the corrections, we focus on the following tensor product decompositions:
\bea \label{tensordecompSO8}
\wedge^3 \fun_{SO(8)} &= \wedge^3 [1,0,0,0] = [0,0,1,1]~, \nn \\
\wedge^4 \fun_{SO(8)} &= \wedge^4 [1,0,0,0] = [0,0,2,0] +[0,0,0,2]~, \nn \\
\text{HWR in}~\fun_{SO(8)} \otimes \wedge^3 \fun_{SO(8)} &= [1,0,1,1]~, \nn\\
\text{HWR in}~\Sym^2  \wedge^3 \fun_{SO(8)} &= [0,0,2,2]~.
%\text{HWR in}~\Sym^2 \Adj_{SO(8)} &= [0,2,0,0]~. 
 \eea
Comparing \eref{tensordecompSO8} with Tables \ref{genuniv2SON}, we see that the corrections to \eref{gen2SON}, when working with $SO(8)$, are originated from $\wedge^3 \fun_{SO(8)}$, $\wedge^4 \fun_{SO(8)}$.  For example, the contributions from $\wedge^4 \fun_{SO(8)}$ are two generators, $[0,0,2,0]$ and $[0,0,0,2]$, of the lattice at order 4.  These are associated with the indices $l_4$ and $m_4$ respectively.

Observe that for the case of $SO(8)$ we have in total 4 generators at order 4, namely $[0;2,0, 0, 0]$, $[0;0,0,2,0]+[0;0,0,0,2]$ and $[0;0,0,0,0]$.  However, from \eref{gen2SON}, there appear only 3 generators at order 4, namely $[0;2,0,0,0]$, $[0;0,0,0,1]$ and $[0;0,0,0,0]$.  Since the dimension of the lattice should be fixed, the presence of an additional generator explains the absence of the order 8 generator, previously associated with the index $n_8$, of the universal lattice.

\begin{figure}[htbp]
\begin{center}
\includegraphics[scale=1]{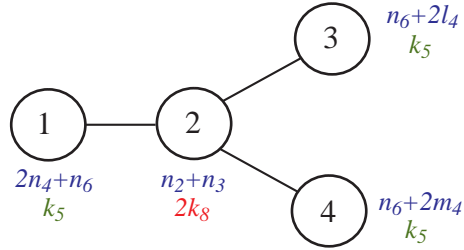}
\caption{The Dynkin diagram of $SO(8)$.  The labels in black indicate ordering of the nodes; the one with number $n$ can be associated with the representation $[0,\ldots, 0, 1, 0, \ldots, 0]$ of $SO(8)$, with $1$ in the $n$-th position from the left. The labels in blue indicate the indices in the universal lattice $f(0;0,\ldots,0)$. The labels in red and green indicate the indices in the non-universal lattices. Observe that the lattice structures are invariant under the triality.}
\label{DynkinD4}
\end{center}
\end{figure}

% and $\Sym^2  \wedge^3 \fun_{SO(8)}$.  

%Note that there is the no order 8 generator, previously associated with the index $n_8$, appearing in the universal lattice.   Such disappearance of such a generator can be seen from the existence of the additional generator $[0,0,2,0]$, associated with the index $l_4$, at order 4 and the fact that the lattice should be 8 dimensional.

The non-universal lattice and the shifts for the case of $SO(8)$ can be determined in a similar way. 
We summarise the relevant information below.

\begin{table}[htdp]
\begin{center}
\small
\begin{tabular}{|c||c|}
\hline
Order of $t$ & $SO(8)$ Dynkin label \\
\hline
2 & $[2;0,0,0]$ and $[0;0,1,0,0]$ \\
\hline
3 & $[1;0,1,0, 0]$ \\
\hline
4 & $[0;2,0, 0, 0]$, $[0;0,0,2,0]+[0;0,0,0,2]$ and $[0;0, \ldots, 0]$  \\
\hline
6 & $[0;1,0,1,1]$ \\
\hline
\end{tabular}
\end{center}
\caption{Generators of the universal lattice for the Hilbert series of 2 $SO(8)$ instantons.}
\label{genuniv2SO8}
\end{table}%

\begin{table}[htdp]
\begin{center}
\small
\begin{tabular}{|c||c|c|}
\hline
Order of $t$ & $SO(8)$ Dynkin label  \\
\hline
5 & $[1;1,0,1,1] $  \\
\hline
8 & $[0;0,2,0,0] $ \\
\hline
\end{tabular}
\end{center}
\caption{Generators of the non-universal lattice for the Hilbert series of 2 $SO(8)$ instantons.}
\label{gennonuniv2SO8}
\end{table}%

\begin{table}[htdp]
\begin{center}
\small
\begin{tabular}{|c||c|c|}
\hline
Order of $t$ & Shift \\
\hline
0 & $[0; 0,0,0, 0] $ \\
\hline
5 (universal lattice) & $[1;0,1,0, 0]$ \\
\hline
5 (non-universal lattice) & $[1;1,0,1,1] $  \\
\hline
7 & $[1;1,0,1,1] $ \\
\hline 
10 & $[0;1,0,1,1]$ \\
\hline
12 & $[2;2,0,2, 2]$  \\
\hline
\end{tabular}
\end{center}
\caption{Shifts from the universal and non-universal lattices at various orders of $t$.}
\label{shifts2SO8}
\end{table}%

Explicitly, the function $f$ for 2 $SO(8)$ instantons is given by
\bea
&f(a;b_1, \ldots, b_r) = \frac{1}{1-t^4} \sum_{m_2 =0}^\infty \sum_{n_2 =0}^\infty\sum_{n_3 =0}^\infty \sum_{n_4 =0}^\infty\sum_{m_4 =0}^\infty \sum_{l_4 =0}^\infty \sum_{n_6 =0}^\infty t^{2 m_2 + 2 n_2 + 3 n_3 + 4 n_4 + 4 m_4 + 4 l_4 + 6 n_6 } \times \nn\\
& [2 m_2 + n_3 + a; 2 n_4 + n_6 +b_1, n_2 + n_3+b_2 , n_6 + 2 l_4+b_3, 2m_4+n_6+b_4] ~,
\eea
and the Hilbert series of the reduced instanton moduli space can be written as
\bea  \label{gen2SO8}
 \widetilde{g}_{2,SO(8)} (t,x,y_1, \ldots, y_4) &=  \sum_{k_8=0}^\infty \Bigg 
\{ f(0;0,2k_8,0,0) t^{8k_8} + f(1;0,2k_8+1,0,0) t^{8k_8+5} + f(1;1,2k_8,1, 1) t^{8k_8+7} \nn \\
& \quad +f(0;1,2k_8+1,1,1) t^{8 k_8+10} \Bigg \} + \sum_{k_5=0}^\infty \Bigg \{ f(k_5+1; k_5+1, 0, k_5+1, k_5+1) t^{5 k_5+5}  \nn\\
& \quad +f(k_5+2; k_5+2, 0, k_5+2, k_5+2) t^{5 k_5+12} \Bigg \} ~.
\eea
%Explicitly, we have 
%
%\bea
%& \widetilde{g}_{2,SO(8)} (t,x,y_1, \ldots, y_4) = \frac{1}{1-t^4} \sum_{m_2 =0}^\infty \sum_{n_2 =0}^\infty\sum_{n_3 =0}^\infty \sum_{n_4 =0}^\infty\sum_{m_4 =0}^\infty \sum_{l_4 =0}^\infty \sum_{n_6 =0}^\infty  \sum_{k_8 =0}^\infty   \nn\\
%& \Bigg\{ [2 m_2 + n_3; 2 n_4 + n_6, n_2 + n_3 + 2 k_8, n_6 + 2 l_4, 2m_4+n_6] t^{2 m_2 + 2 n_2 + 3 n_3 + 4 n_4 + 4 m_4+4 l_4 + 6 n_6 + 8 k_8 } \nn \\
%& + [2 m_2+n_3+1;2 n_4+n_6,n_2+n_3+2 k_8+1,n_6+2 l_4,2m_4+n_6] t^{2 m_2+2 n_2+3 n_3+4 n_4+4 m_4+4 l_4+6 n_6+8 k_8+5} \nn \\
%& +[2 m_2+n_3+1;2 n_4+n_6+1,n_2+n_3+2 k_8,n_6+2 l_4+1,2m_4+n_6+1] t^{2 m_2+2 n_2+3 n_3+4 n_4+4 m_4+4 l_4+6 n_6+8 k_8+7} \nn \\
%& +[2 m_2+n_3;2 n_4+n_6+1,n_2+n_3+2 k_8+1,n_6+2 l_4+1,2m_4+n_6+1] t^{2 m_2+2 n_2+3 n_3+4 n_4+4 m_4+4 l_4+6 n_6+8 k_8+10 } \Bigg\} \nn \\
%& + \frac{1}{1-t^4} \sum_{m_2 =0}^\infty \sum_{n_2 =0}^\infty\sum_{n_3 =0}^\infty \sum_{n_4 =0}^\infty\sum_{m_4 =0}^\infty \sum_{l_4 =0}^\infty  \sum_{k_5 =0}^\infty  \sum_{n_6 =0}^\infty  \nn \\
%& \Bigg \{ [2 m_2+n_3+k_5+1;2 n_4+n_6+k_5+1,n_2+n_3,n_6+2 l_4+k_5+1,2m_4+n_6+k_5+1] t^{2 m_2+2 n_2+3 n_3+4 n_4+4 m_4+4 l_4+6 n_6+5 k_5+5}\nn \\
%& +[2 m_2+n_3+k_5+2;2 n_4+n_6+k_5+2,n_2+n_3,n_6+2 l_4+k_5+2,2m_4+n_6+k_5+2] t^{2 m_2+2 n_2+3 n_3+4 n_4+4 m_4+4 l_4+6 n_6+5 k_5+12} \Bigg \}~.
%\eea
The unrefined Hilbert series is
\bea
 \widetilde{g}_{2,SO(8)} (t,x,\{y_i=1\}) &= \frac{1}{(1-t)^{22} (1+t)^{12} \left(1+t+t^2\right)^{11}} \Big( 1+t+20 t^2+65 t^3+254 t^4+841 t^5+2435 t^6 \nn \\
 & +6116 t^7+14290 t^8+29700 t^9+55947 t^{10}+96519 t^{11}+152749 t^{12}+220408 t^{13} \nn \\
 & +293226 t^{14}+359742 t^{15}+406014 t^{16}+421960 t^{17}+406014 t^{18}+ \text{palindrome up to $t^{34}$} \Big)~.
\eea

Note that out of all theories which have an ADHM construction and correspondingly a weakly coupled UV limit, there is one special theory for which the beta function is zero. This is the theory for the moduli space of $SO(8)$ instantons and is believed to be conformal to all scales. The conformal property allows it to have a supersymmetric index, which in an appropriate limit reduces to the HL index~\cite{Gadde:2011uv,Gaiotto:2012uq}.  The above results are in agreement with the HL index.

\subsubsection{The case of $SO(7)$}
In order to determine the corrections to \eref{gen2SON}, we focus on the following tensor product decompositions:
\bea
\wedge^3 \fun_{SO(7)} &= \wedge^3 [1,0,0] = [0,0,2]~, \nn \\
\wedge^4 \fun_{SO(7)} &= \wedge^4 [1,0,0] = [0,0,2]~, \nn \\
\text{HWR in}~\fun_{SO(7)} \otimes \wedge^3 \fun_{SO(7)} &= [1,0,2]~, \nn\\
%\text{HWR in}~\Sym^2  \wedge^3 \fun_{SO(7)} &= [0,0,4]~, \nn \\
\text{HWR in}~\Sym^2 \Adj_{SO(7)} &= [0,2,0] ~, \label{tensordecompSO7}
 \eea
%From $\fun_{SO(7)} \otimes \wedge^3 \fun_{SO(7)}$, the generator of the universal lattice at order 6, denoted by $n_6$, appearing 
 
\begin{figure}[htbp]
\begin{center}
\includegraphics[scale=0.4]{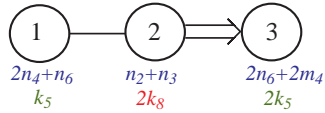}
\caption{The Dynkin diagram of $SO(7)$.  The labels in black indicate ordering of the nodes; the one with number $n$ can be associated with the representation $[0,\ldots, 0, 1, 0, \ldots, 0]$ of $SO(7)$, with $1$ in the $n$-th position from the left. The labels in red and green indicate the indices in the non-universal lattices.}
\label{DynkinB3}
\end{center}
\end{figure} 
 
Explicitly, the function $f$ for 2 $SO(7)$ instantons is given by
\bea
&f(a;b_1, b_2, b_3) = \frac{1}{1-t^4} \sum_{m_2 =0}^\infty \sum_{n_2 =0}^\infty\sum_{n_3 =0}^\infty \sum_{n_4 =0}^\infty\sum_{m_4 =0}^\infty \sum_{n_6 =0}^\infty \nn\\
& [2 m_2 + n_3 + a; 2 n_4 + n_6 +b_1, n_2 + n_3+b_2, 2m_4+2n_6+b_4] t^{2 m_2 + 2 n_2 + 3 n_3 + 4 n_4 + 4 m_4  + 6 n_6  }~,
\eea
and the Hilbert series of the reduced instanton moduli space can be written as
\bea  \label{gen2SO7}
 \widetilde{g}_{2,SO(7)} (t,x,y_1, y_2, y_3) &=  \sum_{k_8=0}^\infty \Bigg 
\{ f(0;0,2k_8,0) t^{8k_8} + f(1;0,2k_8+1,0) t^{8k_8+5} + f(1;1,2k_8,2) t^{8k_8+7} \nn \\
& \quad +f(0;1,2k_8+1,2) t^{8 k_8+10} \Bigg \} + \sum_{k_5=0}^\infty \Bigg \{ f(k_5+1; k_5+1, 0, 2k_5+2) t^{5 k_5+5}  \nn\\
& \quad +f(k_5+2; k_5+2, 0, 2k_5+4) t^{5 k_5+12} \Bigg \} ~.
\eea
 %The Hilbert series can be written as
%
%\bea
%& \widetilde{g}_{2,SO(7)} (t,x,y_1,y_2,y_3) = \frac{1}{1-t^4} \sum_{m_2 =0}^\infty \sum_{n_2 =0}^\infty\sum_{n_3 =0}^\infty \sum_{n_4 =0}^\infty\sum_{m_4 =0}^\infty \sum_{n_6 =0}^\infty  \sum_{k_8 =0}^\infty \nn\\
%& \Bigg\{ [2 m_2 + n_3; 2 n_4 + n_6, n_2 + n_3 + 2 k_8, 2n_6 + 2 m_4] t^{2 m_2 + 2 n_2 + 3 n_3 + 4 n_4 + 4 m_4 + 6 n_6+ 8 k_8} \nn \\
%& + [2 m_2+n_3+1;2 n_4+n_6,n_2+n_3+2 k_8+1,2n_6+2 m_4] t^{2 m_2+2 n_2+3 n_3+4 n_4+4 m_4+6 n_6+8 k_8+5} \nn \\
%& +[2 m_2+n_3+1;2 n_4+n_6+1,n_2+n_3+2 k_8,2n_6+2 m_4+2] t^{2 m_2+2 n_2+3 n_3+4 n_4+4 m_4+6 n_6+8 k_8+7} \nn \\
%& +[2 m_2+n_3;2 n_4+n_6+1,n_2+n_3+2 k_8+1,2n_6+2 m_4+2] t^{2 m_2+2 n_2+3 n_3+4 n_4+4 m_4+6 n_6+8 k_8+10 } \Bigg\} \nn \\
%& + \frac{1}{1-t^4} \sum_{m_2 =0}^\infty \sum_{n_2 =0}^\infty\sum_{n_3 =0}^\infty \sum_{n_4 =0}^\infty\sum_{m_4 =0}^\infty \sum_{k_5 =0}^\infty  \sum_{n_6 =0}^\infty \nn \\
%& \Bigg \{ [2 m_2+n_3+k_5+1;2 n_4+n_6+k_5+1,n_2+n_3,2n_6+2 m_4+2k_5+2] t^{2 m_2+2 n_2+3 n_3+4 n_4+4 m_4+6 n_6+5 k_5+5}\nn \\
%& +[2 m_2+n_3+k_5+2;2 n_4+n_6+k_5+2,n_2+n_3,2n_6+2 m_4+2k_5+4] t^{2 m_2+2 n_2+3 n_3+4 n_4+4 m_4+6 n_6+5 k_5+12} \Bigg \}~. \label{gen2SO7}
%\eea
The unrefined Hilbert series is
\bea
 \widetilde{g}_{2,SO(7)} (t,x,\{y_i=1\}) &=  \frac{1}{(1-t)^{18} (1+t)^{10} (1+t+t^2)^9}  \Big( 1+t+15 t^2+48 t^3+152 t^4+446 t^5+1126 t^6 \nn \\
& \quad +2374 t^7+4674 t^8+8184 t^9+12680 t^{10}+17816 t^{11}+22957 t^{12}+26449 t^{13}  \nn \\
&\quad  +27622 t^{14}+26449 t^{15} + \text{palindrome up to $t^{28}$} \Big)~.
\eea

\subsubsection{The case of $SO(6)$}
Since the Lie algebra of $SO(6)$ is isomorphic to that of $SU(4)$, we expect that the Hilbert series of two $SO(6)$ instantons can be obtained from that of two $SU(4)$ instantons \eref{gen2SU4} with some permutations of the Dynkin labels.  Indeed, 
\bea \label{gen2SO6}
& \widetilde{g}_{2,SO(6)} (t,x,y_1,y_2,y_3)  \nn \\
& =   \sum_{k_4 =0}^\infty   \Bigg \{ f(0; 0,k_4, k_4) t^{ k_4} +f(1; 0,k_4+1, k_4+1) t^{ 4k_4+5} \Bigg \} \nn \\
& + \sum_{k_4 =0}^\infty\sum_{k_6 =0}^\infty   \Bigg \{ \Big[ f(0; 1+k_6,k_4,2+k_4+2 k_6) +f(0; 1+k_6,2+k_4+2 k_6,k_4) \Big] t^{4 k_4 + 6 k_6 + 6} \nn \\
&\qquad \qquad \quad + \Big[ f(1; 1+k_6,k_4,2+k_4+2 k_6) +f(1; 1+k_6,2+k_4+2 k_6,k_4) \Big] t^{4 k_4 + 6 k_6 + 7} \Bigg \}  \nn \\
& +  \sum_{k_5 =0}^\infty \sum_{k_6 =0}^\infty  \Bigg \{ \Big[ f(k_5+1; 1+k_5+k_6,0,2+2 k_5+2 k_6) + f(k_5+1;1+k_5+k_6,2+2 k_5+2 k_6,0) \Big] t^{5 k_5 + 6 k_6 + 5} \nn \\
& \qquad \qquad \quad + \Big[ f(k_5+2; 2+k_5+k_6,0,4+2 k_5+2 k_6) + f(k_5+2; 2+k_5+k_6,4+2 k_5+2 k_6,0) \Big] t^{5 k_5 + 6 k_6 + 12}   \Bigg \}~,
\eea
where the function $f$ is defined as follows:
\bea
 f(a;b_1, b_2,b_3) &= \frac{1}{1-t^4} \sum_{m_2=0}^\infty \sum_{n_2=0}^\infty \sum_{n_3=0}^\infty [2m_2+n_3+a; 2n_4+b_1, n_2+n_3+b_2,   n_2+n_3+b_{3}]  t^{2 m_2+2 n_2+3 n_3}~. \label{deffSO6}
\eea
%The unrefined Hilbert series is equal to 

\section{Two $E_6$ instantons} \label{sec:2E6insts}
\begin{figure}[htbp]
\begin{center}
\includegraphics[scale=0.4]{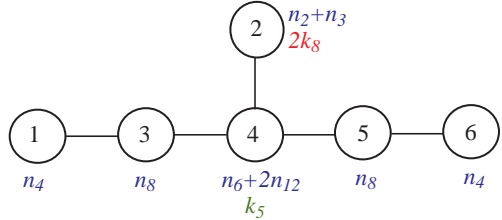}
\caption{The Dynkin diagram of $E_6$.  The labels in black indicate ordering of the nodes; the one with number $n$ can be associated with the representation $[0,\ldots, 0, 1, 0, \ldots, 0]$ of $E_6$, with $1$ in the $n$-th position from the left.   The labels in blue indicate the indices in the universal lattice $f(0;0,\ldots,0)$. The labels in red and green indicate the indices in the non-universal lattices.}
\label{DynkinE6}
\end{center}
\end{figure}

The Hilbert series for two $E_6$ instantons is proposed in (3.13) of \cite{Gaiotto:2012uq} in terms of Hall-Littlewood indices.  
Based on this information, we conjecture that this can be written in terms of an $SU(2) \times E_6$ invariant character expansion as follows:
\bea
\widetilde{g}_{2,E_6} (t,x,y_1, \ldots, y_6) &= \sum_{k_8=0}^\infty  \Big\{ f(0; 0,2k_8,0,0,0,0)t^{8k_8} + f(1; 0,2k_8+1,0,0,0,1)t^{8k_8+5} \nn \\
& \qquad + f(1; 0,2k_8,0,1,0,0)t^{8k_8+7} + f(0; 0, 2k_8+1,0,1,0,0)t^{8k_8+10} \Big \} + \nn\\
& \quad \sum_{k_5=0}^\infty \Big \{ f(k_5+1; 0,0,0,k_5+1,0,0)t^{5k_5+5}+f(k_5+2; 0,0,0,k_5+2,0,0)t^{5k_5+12} \Big \}~, \label{gen2E6}
\eea
where the function $f$ is defined as
\bea
 f(a;b_1, \ldots, b_{6}) &= \frac{1}{1-t^4} \sum_{m_2=0}^\infty \sum_{n_2=0}^\infty\sum_{n_3=0}^\infty \sum_{n_4=0}^\infty \sum_{n_6=0}^\infty \sum_{n_8=0}^\infty \sum_{n_{12}=0}^\infty  t^{2 m_2+2 n_2+3 n_3+4 n_4+6 n_6+8 n_8+12 n_{12}} \times \nn \\
 & \quad [2 m_2+n_3+a ; n_4+b_1,n_2+n_3+b_2,n_8+b_3,n_6+2 n_{12}+b_4,n_8+b_5,n_4+b_6]~.
\eea
The lattice structure is summarised in \fref{DynkinE6}.

As a test of the conjecture, one can check that the character expansion satisfies the limits \eref{limxtgen}, as required.

\section{Two $E_7$ instantons}
The Hilbert series for two $E_7$ instantons is proposed in (A.16) of \cite{Gaiotto:2012uq} in terms of Hall-Littlewood indices.  We conjecture that this can be written in terms of an $SU(2) \times E_7$ invariant character expansion as follows:
\bea
& \widetilde{g}_{2,E_7} (t,x,y_1, \ldots, y_7) \nn \\
&= \sum_{k_8=0}^\infty  \Big\{ f(0; 2k_8,0,0,0,0,0,0)t^{8k_8} + f(1; 2k_8+1,0,0,0,0,0,0)t^{8k_8+5} \nn \\
& \qquad \quad + f(1; 2k_8,0,1,0,0,0,0)t^{8k_8+7} + f(0;  2k_8+1,0,1,0,0,0)t^{8k_8+10} \Big \} + \nn\\
& \sum_{k_5=0}^\infty \Big \{ f(k_5+1; 0,0,k_5+1,0,0,0,0)t^{5k_5+5}+f(k_5+2; 0,0,k_5+2,0,0,0,0)t^{5k_5+12} \Big \}~,
\eea
where the function $f$ is defined as
\bea
 f(a;b_1, \ldots, b_{7}) 
 &= \frac{1}{1-t^4} \sum_{m_2=0}^\infty \sum_{n_2=0}^\infty\sum_{n_3=0}^\infty \sum_{n_4=0}^\infty \sum_{n_6=0}^\infty \sum_{n_8=0}^\infty \sum_{n_{12}=0}^\infty  t^{2 m_2+2 n_2+3 n_3+4 n_4+6 n_6+8 n_8+12 n_{12}} \times \nn \\
 & \quad [2 n_2+n_3+a ; n_2+n_3+b_1, b_2, n_6+2 n_{12}+b_3, n_8+b_4, b_5, n_4+b_6, b_7]~.
\eea
The lattice structures are summarised in \fref{DynkinE7}.  The unrefined Hilbert series of $E_7$ is given in Appendix \ref{sec:E7}

\begin{figure}[htbp]
\begin{center}
\includegraphics[scale=0.5]{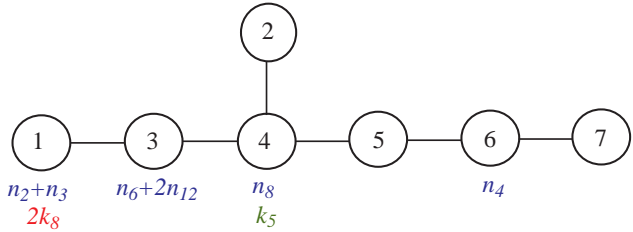}
\caption{The Dynkin diagram of $E_7$.  The labels in black indicate ordering of the nodes; the one with number $n$ can be associated with the representation $[0,\ldots, 0, 1, 0, \ldots, 0]$ of $E_7$, with $1$ in the $n$-th position from the left.   The labels in blue indicate the indices in the universal lattice $f(0;0,\ldots,0)$. The labels in red and green indicate the indices in the non-universal lattices.}
\label{DynkinE7}
\end{center}
\end{figure} 

As a test of the conjecture, one can check that the character expansion satisfies the limits \eref{limxtgen}, as required.

\section{Two $E_8$ instantons}
The Hilbert series is 
\bea
& \widetilde{g}_{2,E_8} (t,x,y_1, \ldots, y_8) \nn \\
&= \sum_{k_8=0}^\infty  \Big\{ f(0; 0,0,0,0,0,0,0,2k_8)t^{8k_8} + f(1; 0,0,0,0,0,0,0,2k_8+1)t^{8k_8+5} \nn \\
& \qquad + f(1; 0,0,0,0,0,0,1,2k_8+1)t^{8k_8+7} + f(0; 0,0,0,0,0,0,1,2k_8+1)t^{8k_8+10} \Big \} + \nn\\
& \quad \sum_{k_5=0}^\infty \Big \{ f(k_5+1; 0,0,0,0,0,0,k_5+1,0)t^{5k_5+5}+f(k_5+2; 0,0,0,0,0,0,k_5+2,0)t^{5k_5+12} \Big \}~,
\eea
where the function $f$ is defined as
\bea
 f(a;b_1, \ldots, b_{8}) &= \frac{1}{1-t^4} \sum_{m_2=0}^\infty \sum_{n_2=0}^\infty\sum_{n_3=0}^\infty \sum_{n_4=0}^\infty \sum_{n_6=0}^\infty \sum_{n_8=0}^\infty \sum_{n_{12}=0}^\infty  t^{2 m_2+2 n_2+3 n_3+4 n_4+6 n_6+8 n_8+12 n_{12}} \times \nn \\
 & \quad [2 n_2+n_3+a ; n_4+b_1,b_2,b_3,b_4,b_5,n_8+b_6,n_6+2 n_{12}+b_7,n_2+n_3+b_8]~.
\eea

\begin{figure}[htbp]
\begin{center}
\includegraphics[scale=0.6]{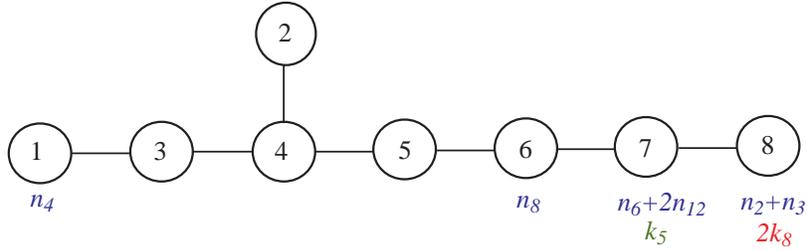}
\caption{The Dynkin diagram of $E_8$.  The labels in black indicate ordering of the nodes; the one with number $n$ can be associated with the representation $[0,\ldots, 0, 1, 0, \ldots, 0]$ of $E_8$, with $1$ in the $n$-th position from the left.   The labels in blue indicate the indices in the universal lattice $f(0;0,\ldots,0)$. The labels in red and green indicate the indices in the non-universal lattices.}
\label{DynkinE8}
\end{center}
\end{figure} 

As a test of the conjecture, one can check that the proposed character expansion satisfies the limits \eref{limxtgen}, as required.

\section{Two $G_2$ instantons}
\begin{figure}[htbp]
\begin{center}
\includegraphics[scale=0.3]{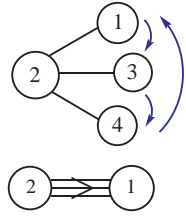}
\caption{Folding the $D_4$ Dynkin diagram to obtain the $G_2$ Dynkin diagram.}
\label{foldingD4toG2}
\end{center}
\end{figure}

%The Hilbert series of two $G_2$ instantons can be obtained from the one of two $SO(8)$ instantons.   
The Dynkin diagram $G_2$ can be obtained by folding the Dynkin diagram of $SO(8)$ via a $\BZ_3$ outer-automorphism, depicted in \fref{foldingD4toG2}.  

We thus have a fugacity map
\bea
y_1' = y_1~,\qquad y_2' = y_2~, \qquad y_3' = y_1~, \qquad y_4' = y_1.
\eea
where $y_1$ and $y_2$ are fugacities for $G_2$ and $y_1', \ldots, y_4'$ are fugacities for $SO(8)$.  Thus, in order to obtain the Hilbert series of two $G_2$ instantons, we simply apply such a map to the representations in Tables \ref{genuniv2SO8}, \ref{gennonuniv2SO8} and \ref{shifts2SO8} and take the highest weight vector after the projection.  Explicitly, we obtain
\bea
[0,1,0,0] &\longrightarrow  [0,1] \nn \\
[2,0,0,0], [0,0,2,0], [0,0,0,2] &\longrightarrow [2,0] \nn \\
[0,0,0,0] &\longrightarrow [0,0] \nn \\
[1,0,1,1] &\longrightarrow [3,0] \nn \\
[2,0,2,2] &\longrightarrow [6,0]~.
\eea

\begin{figure}[htbp]
\begin{center}
\includegraphics[scale=0.8]{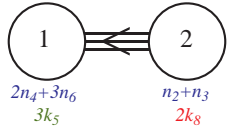}
\caption{The Dynkin diagram of $G_2$.  The labels in black indicate ordering of the nodes; the one with number $n$ can be associated with the representation $[0,\ldots, 0, 1, 0, \ldots, 0]$ of $G_2$, with $1$ in the $n$-th position from the left.   The labels in blue indicate the indices in the universal lattice $f(0;0,\ldots,0)$. The labels in red and green indicate the indices in the non-universal lattices.}
\label{DynkinG2}
\end{center}
\end{figure}

Therefore, we conjecture that he Hilbert series for two $G_2$ instantons is
\bea
& \widetilde{g}_{2,G_2} (t,x,y_1,y_2) \nn \\
&= \sum_{k_8=0}^\infty  \Big\{ f(0; 0,2k_8)t^{8k_8} + f(1; 0,2k_8+1)t^{8k_8+5} + f(1; 3,2k_8)t^{8k_8+7} + f(0;3,2k_8+1)t^{8k_8+10} \Big \} + \nn\\
& \quad \sum_{k_5=0}^\infty \Big \{ f(k_5+1; 3k_5+3,0)t^{5k_5+5}+f(k_5+2; 3k_5+6,0)t^{5k_5+12} \Big \}~,
\eea
where
\bea
f(a;b_1, b_2) & = \frac{1}{1-t^4} \sum_{m_2=0}^\infty \sum_{n_2=0}^\infty \sum_{n_3=0}^\infty  \sum_{n_4=0}^\infty \sum_{n_6=0}^\infty  t^{2 m_2 + 2 n_2 + 3 n_3 + 4 n_4 + 6 n_6} \times \nn \\
& \quad [2 m_2 + n_3 + a; 2 n_4 + 3 n_6 + b_1, n_2 + n_3 +b_2] 
\eea
The lattice structures are summarised in \fref{DynkinG2}.

The unrefined Hilbert series is
\bea
 \widetilde{g}_{2,G_2} (t,x,\{ y_i=1\}) &=\frac{1}{(1-t)^{14} (1+t)^8 (1+t+t^2)^6} \Big( 1+t+10 t^2+31 t^3+75 t^4+180 t^5+385 t^6+637 t^7 \nn \\
&  +975 t^8+1360 t^9+1614 t^{10}+1666 t^{11}+1614 t^{12} + \text{palindrome up to $t^{22}$} \Big)~.
\eea

As a test of the conjecture, one can check that the proposed character expansion satisfies the limits \eref{limxtgen}, as required.

\section{Two $F_4$ instantons} \label{sec:2F4insts}
\begin{figure}[htbp]
\begin{center}
\includegraphics[scale=0.7]{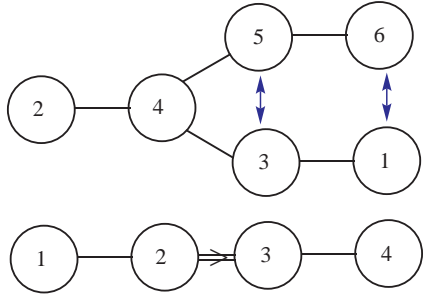}
\caption{Folding the $E_6$ Dynkin diagram to obtain the $F_4$ Dynkin diagram.}
\label{foldingE6toF4}
\end{center}
\end{figure}

The Dynkin diagram $F_4$ can be obtained by folding the Dynkin diagram of $E_6$ via a $\BZ_2$ outer-automorphism, as depicted in \fref{foldingE6toF4}. 

\begin{figure}[htbp]
\begin{center}
\includegraphics[scale=0.4]{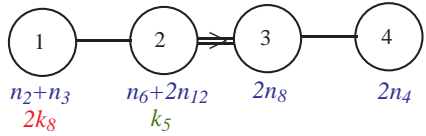}
\caption{The Dynkin diagram of $F_4$.  The labels in black indicate ordering of the nodes; the one with number $n$ can be associated with the representation $[0,\ldots, 0, 1, 0, \ldots, 0]$ of $F_4$, with $1$ in the $n$-th position from the left.   The labels in blue indicate the indices in the universal lattice $f(0;0,\ldots,0)$. The labels in red and green indicate the indices in the non-universal lattices.}
\label{DynkinF4}
\end{center}
\end{figure}

We thus obtain a fugacity map
\bea
y'_1\to y_4, \qquad y'_2\to y_1, \qquad y'_3\to y_3, \qquad y'_4\to y_2, \qquad y'_5\to y_3, \qquad y'_6\to y_4~,
\eea
where $y'_1, \ldots, y'_6$ are fugacities of $E_6$ and $y_1, \ldots, y_4$ are fugacities of $F_4$.
Thus, in order to obtain the Hilbert series of two $G_2$ instantons, we simply apply such a map to the representations in the lattices of \eref{gen2E6} and take the highest weight vector after the projection. 
% Explicitly, we obtain
%\bea
%[0,1,0,0,0,0] &\to  [1,0,0,0] \nn \\
%[1,0,0,0,0,0] &\to [0,0,0,1]\nn \\
%[0,0,0,1,0,0] &\to [0,1,0,0] \nn \\
%\eea
Therefore, we conjecture that the Hilbert series of two $F_4$ instantons is
\bea
 \widetilde{g}_{2,F_4} (t,x,y_1,\ldots,y_4) &= \sum_{k_8=0}^\infty \Big \{f(0;2k_8,0,0,0) t^{8 k_8}+ f(1;2k_8+1,0,0,0) t^{8 k_8+5} + f(1;2k_8,1,0,0) t^{8 k_8+7}  \nn \\
 &+ f(0;2k_8+1,1,0,0) t^{8 k_8+10} \Big \} + \sum_{k_5=0}^\infty  \Big \{ f(k_5+1;0, k_5+1,0,0) t^{5k_5+5} \nn \\
 & + f(k_5+2; 0, k_5+2,0,0) t^{5k_5+12} \Big \} ~.
\eea
where
\bea
f (a;b_1, \ldots, b_4) &= \frac{1}{1-t^4} \sum_{m_2=0}^\infty \sum_{n_2=0}^\infty \sum_{n_3=0}^\infty \sum_{n_4=0}^\infty  \sum_{n_6=0}^\infty \sum_{n_8=0}^\infty \sum_{n_{12}=0}^\infty  t^{2 m_2 + 2 n_2 + 3 n_3 + 4 n_4 + 6 n_6 + 8 n_8 + 12 n_{12}} \times \nn \\
& \quad [2 m_2 + n_3 +a ; n_2 + n_3 +b_1, n_6 + 2 n_{12} +b_2, 2 n_8+b_3, 2 n_4 +b_4] 
\eea
The lattice structures are summarised in \fref{DynkinF4}.

The unrefined Hilbert series is given by
\bea
& \widetilde{g}_{2,F_4} (t,x,\{ y_i=1\}) = \frac{1}{(1-t)^{34}(1+t)^{22}(1+t+t^2)^{17}} \times \nn \\
& \Big( 1+5 t+48 t^2+287 t^3+1560 t^4+7503 t^5+32316 t^6+125355 t^7+444325 t^8+1443572 t^9+4322993 t^{10} \nn \\
& +11989241 t^{11}+30913094 t^{12}+74321701 t^{13}+167106519 t^{14}+352245510 t^{15}+697557618 t^{16}+1300152932 t^{17} \nn \\
& +2284606168 t^{18}+3790004228 t^{19}+5943020899 t^{20}+8818128233 t^{21}+12392104012 t^{22}+16505926853 t^{23} \nn \\
& +20851379873 t^{24}+24994963144 t^{25}+28442119825 t^{26}+30731161887 t^{27}+31533797982 t^{28} \nn \\
&+30731161887 t^{29} + \text{palindrome up to $t^{56}$} \Big)~.
\eea

As a test of the conjecture, one can check that the proposed character expansion satisfies the limits \eref{limxtgen}, as required.

\section{Universal features of lattices} \label{sec:universal}
In this section, we discuss certain universal features of lattices that appear in all character expansions of generic instanton gauge groups $G$.  For $G$ being $SU(N)$, $SO(N)$, $Sp(N)$, the term `generic' means the results are correct for all $N \geq N_0$, for some $N_0 >0$.

\subsection{Generators of the lattices} 
For a given simple group $G$, there are precisely two sets of generators of two $G$ instanton moduli space.  The ones at order $t^2$ transform under the $SU(2) \times G$ representation $[2; 0] + [0;\Adj]$, and the ones at order $t^3$ transform under $[1;\Adj]$.  These highest weight vectors are also generators of the universal lattice; the corresponding indices are denoted by $m_2$, $n_2$ and $n_3$ respectively.  Generators at higher orders of $t$ can be extracted from the general formulae. We summarise the generators of the universal and non-universal lattices in \tref{genlatt}.

\begin{table}[htdp]
\begin{center}
\small
\begin{tabular}{|c|c|c|}
\hline
Order of $t$ & Index for lattice generator & Representation of group $G$  \\
\hline
$2$,$3$ & $n_2$, $n_3$ & $\Adj$  \\
\hline
$4$ & $n_4$ & HWR in $\Sym^2 \Adj$  \\
\hline
$4$ & $m_4$ & 2nd HWR in $\Sym^2 \Adj$  \\
\hline
$4$ & $k_4$ & 3rd HWR in  $\Sym^2 \Adj$ \\
\hline
6 &$n_6$, $k_5$,$k_6$ & HWR in $\wedge^2 \Adj$  \\
\hline
8 &$n_8$ & A certain rep. in $\Sym^2 \Adj$  \\
\hline
8 & $k_8$ & HWR in $\Sym^2 \Adj$  \\
\hline
12 & $n_{12}$ & HWR in $\Sym^2 \wedge^2 \Adj$ \\
\hline
\end{tabular}
\end{center}
\caption{Generators for the universal and non-universal lattices in character expansions of generic instanton gauge groups.  The indices $n$'s and $m$'s are associated with the universal lattice, and the indices $k$'s are associated with the non-universal lattice. The acronym `HWR' stands for the highest weight representation.}
\label{genlatt}
\end{table}%

\subsection{Dimension of the lattice and dimension of the moduli space}  
In this subsection, the dimension of the lattice in the character expansion is related to the dimension of the moduli space.  

In a character expansion, the summands contain representations of $SU(2) \times G$.  For each $G$, the general form of representations that appears in the character expansion can be determined from the nodes of Dynkin diagram occupied by the lattices. We tabulate those general forms in the second column of \tref{deltaR}.  Note that the dimensions of such representations are polynomials of $n$.  Let $\delta_R$ be the degree of such a polynomial.  The value of $\delta_R$ for each group is tabulated in \tref{deltaR}. Observe that $\delta_R$ depends on group $G$ and on the number of nodes $N_0$ of the corresponding Dynkin diagram are occupied by the lattices.  
\begin{table}[htdp]
\begin{center}
\scriptsize
\begin{tabular}{|c|c|c|c|c|c|c|c|}
\hline
$G$ & $r_G$& $h^\vee_G$ & \# nodes $N_0$   & Representation & Degree $\delta_R$ of polynomial& Degree $\delta_R$ of polynomial \\
     &  &   & occupied & $\mathbf{R}_n$ & of dimension of $\mathbf{R}_n$  & of dimension of $\mathbf{R}_n$  \\
& & & & &when $r_G \leq N_0$ & when $r_G \geq N_0$\\
\hline
$SU(N)$ & $N-1$ & $N$ & $4$ & $[n,n,0, \ldots, 0,n,n]$ & $\frac{1}{2} (\dim G- r_G) = {N \choose 2}$& $4 h^\vee_{SU(N)} -10$ \\
\hline
$Sp(N)$ & $N$ & $N+1$ & $2$ & $[n,n,0,\ldots,0]$& $\frac{1}{2} (\dim G- r_G) = N^2$ & $4h^\vee_{Sp(N)} -8$ \\
\hline
$SO(N)$ & $\lfloor N/2 \rfloor$  & $N-2$ & $4$ & $[n,n,n,n,0,\ldots,0]$ &  $\frac{1}{2} (\dim G- r_G) $ &  $4h^\vee_{SO(N)} -12$ \\
\hline
$E_6$ & $6$ & 12 & $6$ & $[n,n,n,n,n,n]$ & $\frac{1}{2} (\dim G- r_G) =36$  & $4h^\vee_{E_6} -12$ \\
\hline
$E_7$ & 7 & 18 & $4$ & $[n,0,n,n,0,n,0]$  &-& $4h^\vee_{E_7} -12$\\
\hline
$E_8$ & 8 & 30 & $4$  & $[n,0,0,0,0,n,n,n]$  & -& $4h^\vee_{E_8} -12$ \\
\hline
$G_2$ & 2 & 4& $2$ & $[n,n]$  & $\frac{1}{2} (\dim G- r_G) =6$  & $4h^\vee_{G_2} -10$\\
\hline
$F_4$ & 4 & 9 & $4$ & $[n,n,n,n]$& $\frac{1}{2} (\dim G- r_G) =24$ & $4h^\vee_{F_4} -12$ \\
\hline
\end{tabular}
\end{center}
\caption{In a character expansion, the summands contain representations of $SU(2) \times G$.  Here $\mathbf{R}_n$ denotes a general form of representations for each group, and $\delta_R$ denotes the degree of the polynomial in the dimension of $\mathbf{R}_n$.  The notation $r_G$ and $h_G^\vee$ denote the rank and the dual coxeter number of the group $G$. Note that $\frac{1}{2} (\dim G- r_G)$ is the number of positive roots in the root system of $G$.}
\label{deltaR}
\end{table}%

%\bea
%\dim [n,n,0, \ldots, 0,n,n]_{SU(N)} &\sim n^{4N-10}~, \nn\\
%\dim [n,n,0,\ldots,0]_{Sp(N)} &\sim n^{4N-4}~, \nn \\
%\dim [n,n,n,n,0,\ldots,0]_{SO(N)} &\sim n^{4N-20}~, \nn \\
%\dim [n,n,n,n,n,n]_{E_6} &\sim n^{36}~, \nn \\
%\dim [n,0,n,n,0,n,0]_{E_7} &\sim n^{60}~, \nn \\
%\dim [n,0,0,0,0,n,n,n]_{E_8} &\sim n^{108}~, \nn \\
%\dim [n,n]_{G_2} &\sim n^6~, \nn \\
%\dim [n,n,n,n]_{F_4} &\sim n^{24}~.
%\eea
%The value of $\delta_R$ for each group is simply the power of $n$ listed above; these are tabulated in the third column of \tref{dimlattices}.

\paragraph{Order of the pole at $t=1$.}  Let us {\it estimate} that dimension of the moduli space from the dimension of lattice.  We determine the former from the the order of the pole at $t=1$.   There are three useful observations that can be applied to this question: 
\ben
\item The sum $\sum_{n=0}^\infty a_m(n) t^n$ of a polynomial $a_m(n)$ of degree $D$ is a rational function in $t$ with a pole at $t=1$ of order $D+1$. For example, $\dim [n]_{SU(2)}= n+1$ is of degree 1 and
\bea \label{contrSU2a} \sum_{n=0}^\infty \dim [n]_{SU(2)} t^n = \frac{1}{(1-t)^2}~, \eea
which has a pole at $t=1$ of order $2$.  
\item Each further summation over $t^n$ contributes one order of the pole at $t=1$. 
\item  In addition to the contributions from representations of $G$, we has to take into account the contribution from $SU(2)$.  
Consider $\sum_{n=0}^\infty \dim [n; \mathbf{R}_n] t^n$, where $\mathbf{R}_n$ is a representation listed in the second column of \tref{deltaR}. 
In this way the contribution from $SU(2)$ increases the order of the pole at $t=1$ by $1$.
\een

Let $\delta_{L}$ be the dimension of the lattice present in the character expansion for each group. 
From the above discussion, the relations between the dimension of the reduced two $G$-instanton moduli space and the dimension of the lattice are given by
\bea
\dim_{\BC} \widetilde{\CM}_{2, G} =  4 h^\vee_G-2 = \delta_R + \delta_{L} +1~,
\eea  
where $\delta_R$ are given by \tref{deltaR}.
Thus, the dimension of the lattice can be written as
\bea
\delta_{L} = 4 h^\vee_G -\delta_R -3   ~.
\eea
Let us tabulate the dimension of lattice when the rank of the group is greater than or equal to the number of nodes occupied, $N_0$, in \tref{dimlattices}.

\begin{table}[htdp]
\begin{center}
\small
\begin{tabular}{|c|c|}
\hline
$G$  & $\delta_{L}$\\
\hline
$SU(N)$, $N\geq 4$  & $7$\\
\hline
\hline
$Sp(N)$, $N \geq 2$  & $5$  \\
\hline
\hline
$SO(N)$, $N \geq 8$   & $9$ \\
\hline
$E_6$ & $9$ \\
\hline
$E_7$ & $9$ \\
\hline
$E_8$  & $9$ \\
\hline
$G_2$  & $7$\\
\hline
$F_4$  & $9$\\
\hline

%\hline
%$G$ & $h^\vee$ & \# nodes occupied & $\delta_R$ & $\delta_{L}$\\
%\hline
%$SU(N)$ & $N$ & $4$ & $4 h^\vee_{SU(N)} -10$  & $7$\\
%\hline
%\hline
%$Sp(N)$ & $N+1$& $2$ & $4h^\vee_{Sp(N)} -8$  & $5$  \\
%\hline
%\hline
%$SO(N)$ & $N-2$ & $4$ & $4h^\vee_{SO(N)} -12$  & $9$ \\
%\hline
%$E_6$ & 12& $6$ &$4h^\vee_{E_6} -12$  & $9$ \\
%\hline
%$E_7$ & 18 & $4$ & $4h^\vee_{E_7} -12$  & $9$ \\
%\hline
%$E_8$ & 30 & $4$ & $4h^\vee_{E_8} -12$ & $9$ \\
%\hline
%$G_2$ & 4 & $2$ &$4h^\vee_{G_2} -10$ & $7$\\
%\hline
%$F_4$ & 9 & $4$ & $4h^\vee_{F_4} -12$  & $9$\\
%\hline
\end{tabular}
\end{center}
\caption{The dimension of the lattice, $\delta_L$, for each group. It should be emphasised that, for $SU(N)$, $SO(N)$ and $Sp(N)$, the values of $\delta_L$ listed here are valid when $N$ {\it larger than or equal} to the number of nodes occupied $N_0$.  This is according to what we call `generic', defined at the beginning of the section.}
\label{dimlattices}
\end{table}%

\subsubsection{Higher instanton numbers} 
Let us propose some conjectures for higher instanton numbers.

We first discuss $SU(N)$.  When all nodes in the Dynkin diagram are occupied by the lattice, from \tref{deltaR}, the dimension of the representation $[n,n,\ldots, n,n]$ has a degree ${N \choose 2}$; this is the value of $\delta_R$ when all nodes are occupied.  Thus, in this situation, the dimension of the lattice is given by $\delta_L = 2kN - {N \choose 2} - 3$.  Observe that the dimension of the lattice $\delta_L$ has a critical value at $N=2k+ \frac{1}{2}$, but since $N$ is an integer, $\delta_L$ attains its maximum at $N=2k$.  When $N > 2k$, we predict that the dimension of the lattice should remain at $\delta_L(N=2k) = (2k+3)(k-1)$.  

In conclusion, there is a critical value at $N=2k$; at this value, all nodes in the Dynkin diagram are occupied by the lattice.  For $SU(N)$,
\bea
\delta_L = \begin{cases} {2k+1 \choose 2} - 3 = (2k+3)(k-1)  & \quad \text{for $N > 2k$} \\
2kN - {N \choose 2} - 3 &  \quad \text{for $N \leq 2k$}~.  \end{cases}
\eea
Observe that for $2k<N$, $\delta_L$ is independent of $N$. 

This generalizes for any group $G$. Let $r_G$ be the rank and $h^\vee_G$ be the dual coxeter number.
We conjecture that there is a critical value at $r_G=2k$.
For higher $k$, we conjecture that
\bea
\delta_L = 2 k h^\vee_G - \frac{1}{2} (\dim G- r_G) - 3~, \qquad \text{for $2k \geq r_G$}~.
\eea
For $r_G>2k$, the formula is conjectured to be quadratic in $k$ 

\section{Summary and Outlook}

Let us briefly summarize and discuss our results. In this paper we have discussed in complete detail
the Hilbert series for the moduli space of two instantons of simple groups. Our strategy was to study the cases 
where the instanton moduli space can be identified with the Higgs branch of ${\cal N}=2$ supersymmetric gauge theories
and write the Hilbert series in a way that will be very suggestive for generalizations to other cases. In particular we have 
studied the expressions for the Hilbert series of ABCD cases coming from supersymmetric theories with Lagrangians, {\it i.e.}  ADHM 
construction, and for E cases coming from theories without Lagrangians computed as Hall-Littlewood index.
We have then employed the very restrictive form of this expression as a character sum generated by a very small set of representations
to suggest explicit expressions for the Hilbert series of FG cases. These expressions pass several non-trivial consistency checks.

Let us list several research directions for generalizations, extensions and applications of our results. 
Having obtained the explicit form of the Hilbert series for one~\cite{Benvenuti:2010pq} and two-instanton
moduli space of simple groups in a rather simple form, it is natural to ask whether the Hilbert series for higher
instantons can be explicitly computed. For exceptional groups of type E one can use the Hall-Littlewood index
expressions~\cite{Gaiotto:2012uq}. However, since the results using the index are not packaged in representations of the 
symmetry group but rather in representations of a maximal sub-group\footnote{
For example the HL index for higher rank theories with $E_6$ flavor symmetry is given explicitly
by the  procedure of~\cite{Gaiotto:2012uq} in $SU(3)\times SU(3)\times  SU(2)\times U(1)\times U(1)$ covariant form.} 
these are a-priori hard to generalize to other groups. In particular it would be useful to understand better
the relation of $A_{N-1}$ Hall-Littlewood polynomials, appearing in the expressions for the  index,
to the representation theory of the E-type groups.
Another approach to tackle this problem would be to gain a better understanding of the relatively simple 
character expansion, and in particular the lattice structure we have discussed in detail, the two-instanton Hilbert series
 admits. It might be interesting to see whether this structure is visible in the recursive procedure to generate the Hilbert series
of multi-instantons suggested in~\cite{Nakajima:2003pg}.

Given the explicit expressions of this paper it is also natural to ask what kind of physical information about the instantons
can be extracted from them. We have used the very simple physical property that in certain limits of the moduli space
the interactions between the instantons are negligible to perform consistency checks of the expressions. 
This amounted to computing residues at a certain class of poles of the Hilbert series.
A much more interesting
question would be to extract information about the interactions between the instantons. It might be 
useful here to study the analytic properties, and their physical interpretation, of the Hilbert series in more detail.

%%%%%%%%%%%%%%%%%%%%%%%%%%%

\acknowledgments
We thank Abhijit Gadde, Davide Gaiotto, Christoph Keller, Leonardo Rastelli, Jaewon Song, Yuji Tachikawa, Giuseppe Torri and Alberto Zaffaroni for discussions and correspondence.  We acknowledge Christoph Keller and Jaewon Song for sharing their results with us prior to the publication. We are particularly indebted to Yuji Tachikawa for clarifying issues on computing Hilbert series for theories with orthogonal gauge groups.  

N.~M. would like to express his gratitude towards the following institutes and collaborators during the completion of this project: Kavli Institute for the Physics and Mathematics of the Universe, Progress in Quantum Field Theory and String Theory Workshop (Osaka City University), Maths of String and Gauge Theory Workshop (City University London and King's College London), Mathematical Institute Oxford, Oriel College Oxford, Imperial College London; Yuji Tachikawa, James Sparks, Theerasak Mingarcha, Alexander Shannon and Aroonroj Mekareeya.  The work of N.~M. is supported by a research grant of the Max Planck Society, World Premier International Research Center Initiative, MEXT, Japan, and Winton Capital Prize (awarded by Winton Capital Management).

The research of S.~S.~R. is supported in part by NSF grant PHY-0969448.

%%%%%%%%%%%%%%%%%%%%%%%%%%%
\appendix
\section{The unrefined Hilbert series of reduced two $E_6$ instanton moduli space} \label{sec:E6}
The unrefined Hilbert series of reduced two $E_6$ instanton moduli space can be written as
\bea
\widetilde{g}_{2,E_6} (t,x=1,\{ y_i =1 \}) = \frac{P(t)}{Q(t)}~, 
\eea
where
\bea
Q(t) &= (1-t)^{46} (1+t)^{32} \left(1+t+t^2\right)^{23}~, \nn \\
P(t) &= 1+9 t+94 t^2+739 t^3+5121 t^4+31432 t^5+173895 t^6+874485 t^7+4036298 t^8+17200367 t^9 \nn \\
& +68039474 t^{10}+250943933 t^{11}+866242068 t^{12}+2807705547 t^{13}+8569454706 t^{14}+24690503239 t^{15} \nn \\
& +67304396959 t^{16}+173919980352 t^{17}+426790882149 t^{18}+996158535441 t^{19}+2214670938701 t^{20} \nn \\
& +4695878015170 t^{21}+9507297417908 t^{22}+18398716114730 t^{23}+34066083855696 t^{24}+60399840583490 t^{25} \nn \\
& +102628223553496 t^{26}+167232472484542 t^{27}+261500117384417 t^{28}+392614934492341 t^{29} \nn \\
& +566271723784347 t^{30}+784947220008032 t^{31}+1046126546231772 t^{32}+1340924322289616 t^{33} \nn \\
& +1653587141756229 t^{34}+1962268356880815 t^{35}+2241216639463322 t^{36}+2464163123099051 t^{37} \nn \\
& +2608327634962043 t^{38}+2658213934310966 t^{39}+2608327634962043 t^{40}+ \text{palindrome up to $t^{78}$}~.
\eea

\section{The unrefined Hilbert series of reduced two $E_7$ instanton moduli space} \label{sec:E7}
The unrefined Hilbert series of reduced two $E_7$ instanton moduli space can be written as
\bea
\widetilde{g}_{2,E_7} (t,x=1,\{ y_i =1 \}) = \frac{P(t)}{Q(t)}~, 
\eea
where
{\footnotesize
\bea
Q(t) &= (1-t)^{70} (1+t)^{52} \left(1+t+t^2\right)^{35}~, \nn \\
P(t) &= 1+17 t+237 t^2+2628 t^3+25193 t^4+213819 t^5+1638666 t^6+11476871 t^7+74152233 t^8 \nn \\
& +445070980 t^9+2495671432 t^{10}+13133928036 t^{11}+65121712327 t^{12}+305215505275 t^{13} \nn \\
& +1356033968529 t^{14}+5725284334978 t^{15}+23021851542594 t^{16}+88338636956104 t^{17} \nn \\
& +324035139906700 t^{18}+1138031848052668 t^{19}+3832341391241046 t^{20}+12390621413785440 t^{21} \nn \\
& +38509222288582663 t^{22}+115175603408208175 t^{23}+331836472263902521 t^{24}+921861932483495244 t^{25} \nn \\
& +2471530433876763846 t^{26}+6399961693050532054 t^{27}+16018745367471142680 t^{28} \nn \\
& +38781560068496818142 t^{29}+90876821066275028695 t^{30}+206242719899419463791 t^{31} \nn \\
& +453576963793872584712 t^{32}+967171231109021529977 t^{33}+2000571291562232590513 t^{34} \nn \\
& +4016126507767354504238 t^{35}+7828073649219480743672 t^{36}+14820947289312246349740 t^{37} \nn \\
& +27267076918737091016348 t^{38}+48764087264312469202730 t^{39}+84802326792798968389732 t^{40} \nn \\
& +143449590902653729399624 t^{41}+236104043071240448693797 t^{42}+378216261606533139497461 t^{43} \nn \\
& +589822792928957883073617 t^{44}+895677339869346647226824 t^{45}+1324728639658651633703727 t^{46}\nn \\
& +1908697079658876873038411 t^{47}+2679565476854052143878502 t^{48}+3665936157860425562998541 t^{49} \nn \\
& +4888414479465062757831170 t^{50}+6354435158683924634396271 t^{51}+8053206553397859455383003 t^{52} \nn \\
& +9951646269406905770095206 t^{53}+11992251412402642586454948 t^{54}+14093734406768042617860546 t^{55} \nn \\
& +16154939755233169917249815 t^{56}+18062065264884658609927825 t^{57}+19698620890606501833935055 t^{58} \nn \\
& +20956986683280640928389866 t^{59}+21750009714684524653667914 t^{60}+22020920210850484561094012 t^{61} \nn \\
& +21750009714684524653667914 t^{62}+ \text{palindrome up to $t^{122}$}~.
\eea}

\section{The unrefined Hilbert series of reduced two $E_8$ instanton moduli space} \label{sec:E8}
The unrefined Hilbert series of reduced two $E_8$ instanton moduli space can be written as
\bea
\widetilde{g}_{2,E_8} (t,x=1,\{ y_i =1 \}) = \frac{P(t)}{Q(t)}~, 
\eea
where the numerator $P(t)$ is very lengthy and so we present only the partial result here:
\bea
Q(t) &= (1-t)^{118} (1+t)^{92} (1+t+t^2)^{59}~, \nn \\ 
P(t) &= 1+33 t+720 t^2+12229 t^3+175244 t^4+2202671 t^5+24878783 t^6+256569378 t^7+2443540988 t^8 \nn\\ 
& +21674827640 t^9 +180246622201 t^{10}+1412639547809 t^{11}+10478772278218 t^{12}+73834160727443 t^{13} \nn \\
& +495666418105592 t^{14}+3178631239205683 t^{15}+19516241282632383 t^{16}+114954513088177804 t^{17} \nn \\
& +650730074024913955 t^{18}+3545773840019047915 t^{19}+18624246310032133812 t^{20}  + \ldots  \nn \\
&+ a t^{104} + b t^{105} + a t^{106} + \text{palindrome up to $t^{210}$}~,
 \eea
 with
 \bea
 a &= 1901988330904941153099303084572300620676101660 ~, \nn \\
 b &= 1915944403166935974529194136345112597020458466~.
 \eea

\bibliographystyle{ytphys}
\bibliography{ref}

\end{document}